\documentclass[aps,prd,nofootinbib,twocolumn,superscriptaddress,floatfix,notitlepage]{revtex4-2}

\makeatletter
\newcommand{\fmarki}{$\alpha$}
\newcommand{\fmarkii}{$\beta$}
\newcommand{\fmarkiii}{$\gamma$}
\newcommand{\fmarkiv}{$\delta$}
\newcommand{\fmarkv}{$\epsilon$}
\newcommand{\fmarkvi}{$\zeta$}
\newcommand{\fmarkvii}{$\eta$}
\newcommand{\fmarkviii}{$\theta$}
\newcommand{\fmarkix}{$\iota$}
\newcommand{\fmarkx}{$\kappa$}
                
\def\@fnsymbol#1{{\ifcase#1\or \fmarki\or \fmarkii\or \fmarkiii\or \fmarkiv\or \fmarkv\or \fmarkvi\or \fmarkvii\or \fmarkviii\or \fmarkix\or \fmarkx \else\@ctrerr\fi}}
\makeatother

\usepackage{ulem}
\usepackage{graphicx}
\usepackage{amsmath,amsfonts,amssymb}
\usepackage{color}
\usepackage{verbatim}
\usepackage{enumitem}

\usepackage[table]{xcolor}
\usepackage{natbib}
\usepackage[breaklinks,colorlinks,urlcolor=blue,citecolor=blue,linkcolor=magenta]{hyperref}

\usepackage{latexsym,float,url,mathrsfs}
\usepackage{epstopdf}
\usepackage{comment}
\usepackage{subfigure}
\usepackage{tabularx}
\usepackage{orcidlink}

\usepackage{aas_macros}

\newcommand{\be}{\begin{equation}} 
\newcommand{\ee}{\end{equation}}
\newcommand{\bea}{\begin{equation}\begin{aligned}} 
\newcommand{\eea}{\end{aligned}\end{equation}}
\newcommand{\td}{{\rm d}}
\newcommand{\cM}{\mathcal{M}}
\newcommand{\Msun}{M_\odot}

\definecolor{rossocorsa}{rgb}{0.83, 0.0, 0.0}

\def\lsim{\mathrel{\raise.3ex\hbox{$<$\kern-.75em\lower1ex\hbox{$\sim$}}}}
\def\gsim{\mathrel{\raise.3ex\hbox{$>$\kern-.75em\lower1ex\hbox{$\sim$}}}}

\newcommand{\papertitle}{What is the source of the PTA GW signal?}

\begin{document}
\title[]{\papertitle}

\author{John Ellis\orcidlink{0000-0002-7399-0813}}
\email{john.ellis@cern.ch}
\affiliation{Keemilise ja Bioloogilise F\"u\"usika Instituut, R\"avala pst. 10, 10143 Tallinn, Estonia}
\affiliation{Physics Department, King’s College London, Strand, London, WC2R 2LS, United Kingdom}
\affiliation{Theoretical Physics Department, CERN, CH 1211 Geneva, Switzerland}

\author{Malcolm Fairbairn\orcidlink{0000-0002-0566-4127}}
\email{malcolm.fairbairn@kcl.ac.uk}
\affiliation{Physics Department, King’s College London, Strand, London, WC2R 2LS, United Kingdom}

\author{Gabriele Franciolini\orcidlink{0000-0002-6892-9145}}
\email{gabriele.franciolini@uniroma1.it}
\affiliation{Dipartimento di Fisica, ``Sapienza'' Universit\`a di Roma, Piazzale Aldo Moro 5, 00185, Roma, Italy}
\affiliation{INFN sezione di Roma, Piazzale Aldo Moro 5, 00185, Roma, Italy}

\author{Gert H\"utsi}
\email{gert.hutsi@kbfi.ee}
\affiliation{Keemilise ja Bioloogilise F\"u\"usika Instituut, R\"avala pst. 10, 10143 Tallinn, Estonia}
\author{Antonio~Iovino~\orcidlink{0000-0002-8531-5962}}
\email{antoniojunior.iovino@uniroma1.it}
\affiliation{Dipartimento di Fisica, ``Sapienza'' Universit\`a di Roma, Piazzale Aldo Moro 5, 00185, Roma, Italy}
\affiliation{INFN sezione di Roma, Piazzale Aldo Moro 5, 00185, Roma, Italy}
\affiliation{Keemilise ja Bioloogilise F\"u\"usika Instituut, R\"avala pst. 10, 10143 Tallinn, Estonia}

\author{Marek Lewicki\orcidlink{0000-0002-8378-0107}}
\email{marek.lewicki@fuw.edu.pl}
\affiliation{Faculty of Physics, University of Warsaw ul. Pasteura 5, 02-093 Warsaw, Poland}

\author{Martti~Raidal\orcidlink{0000-0001-7040-9491}}
\email{martti.raidal@cern.ch}
\affiliation{Keemilise ja Bioloogilise F\"u\"usika Instituut, R\"avala pst. 10, 10143 Tallinn, Estonia}

\author{Juan Urrutia\orcidlink{0000-0002-6035-6610}}
\email{juan.urrutia@kbfi.ee}
\affiliation{Keemilise ja Bioloogilise F\"u\"usika Instituut, R\"avala pst. 10, 10143 Tallinn, Estonia}
\affiliation{Departament of Cybernetics, Tallinn University of Technology, Akadeemia tee 21, 12618 Tallinn, Estonia}

\author{Ville Vaskonen\orcidlink{0000-0003-0003-2259}}
\email{ville.vaskonen@pd.infn.it}
\affiliation{Keemilise ja Bioloogilise F\"u\"usika Instituut, R\"avala pst. 10, 10143 Tallinn, Estonia}
\affiliation{Dipartimento di Fisica e Astronomia, Universit\`a degli Studi di Padova, Via Marzolo 8, 35131 Padova, Italy}
\affiliation{INFN sezione di Padova, Via Marzolo 8, 35131 Padova, Italy}

\author{Hardi Veerm\"ae\orcidlink{0000-0003-1845-1355}}
\email{hardi.veermae@cern.ch}
\affiliation{Keemilise ja Bioloogilise F\"u\"usika Instituut, R\"avala pst. 10, 10143 Tallinn, Estonia}

\begin{abstract}
The most conservative interpretation of the nHz stochastic gravitational wave background (SGWB) discovered by NANOGrav and other Pulsar Timing Array (PTA) Collaborations is astrophysical, namely that it originates from supermassive black hole (SMBH) binaries. However, alternative cosmological models have been proposed, including cosmic strings, phase transitions, domain walls, primordial fluctuations and ``audible" axions. We perform a multi-model analysis (MMA) to compare how well these different hypotheses fit the NANOGrav data, both in isolation and in combination with SMBH binaries, and address the questions: Which interpretations fit the data best, and which are disfavoured? We also discuss experimental signatures that can help discriminate between different sources of the PTA GW signal, including fluctuations in the signal strength between frequency bins, individual sources, and how the PTA signal extends to higher frequencies.\\
~~\\
KCL-PH-TH/2023-43, CERN-TH-2023-153, AION-REPORT/2023-08
\end{abstract}

\maketitle

\section{Introduction} 

The Pulsar Timing Array (PTA) collaborations NANOGrav~\cite{NANOGrav:2023gor,NANOGrav:2023hde}, EPTA (in combination with InPTA)\,\cite{EPTA:2023fyk,EPTA:2023sfo,EPTA:2023xxk}, PPTA\,\cite{Reardon:2023gzh,Zic:2023gta,Reardon:2023zen} and CPTA\,\cite{Xu:2023wog} have recently announced the discovery of a stochastic gravitational wave background (SGWB) at frequencies in the nHz range, with a strength and frequency dependence similar to that expected from a population of supermassive black holes (SMBH) binary systems. This discovery opens new perspectives in the studies of GW sources, complementing the mergers of stellar-mass black holes discovered by the LIGO-Virgo-KAGRA collaborations~\cite{LIGOScientific:2018mvr,LIGOScientific:2020ibl,LIGOScientific:2021djp}. Measurements of the nHz SGWB are at an early stage, and it is important to verify to what extent it is consistent with the SMBH binary model. Indeed, the first measurements seem to indicate that the frequency dependence of the SGWB observed by the PTAs may differ somewhat from that predicted in the simplest version of the SMBH binary model, according to which the binaries evolve by losing energy primarily through GW emission\cite{NANOGrav:2023hfp,EPTA:2023xxk}. 

Two classes of interpretation of this apparent discrepancy have been proposed. One is that the SMBH binaries may also be losing energy through some other mechanism, presumably through interactions with their environments, that would reduce the period over which they emit GWs in the frequency range measured \cite{NANOGrav:2023hfp,Ellis:2023dgf,Ghoshal:2023fhh,Shen:2023pan,Broadhurst:2023tus,Bi:2023tib,Zhang:2023lzt}. A more radical interpretation is that the GWs  are being emitted by some cosmological source whose origin is in fundamental physics. Candidate sources that have been considered include a network of cosmic (super)strings\,\cite{Ellis:2023tsl,Kitajima:2023vre,Wang:2023len,Lazarides:2023ksx,Eichhorn:2023gat,Chowdhury:2023opo,Servant:2023mwt,Antusch:2023zjk,Yamada:2023thl,Ge:2023rce,Basilakos:2023xof}, a first-order phase transition in the early Universe\,\cite{Fujikura:2023lkn,Addazi:2023jvg,Bai:2023cqj,Megias:2023kiy,Han:2023olf,Zu:2023olm,Megias:2023kiy,Ghosh:2023aum,Xiao:2023dbb,Li:2023bxy,DiBari:2023upq,Cruz:2023lnq,Gouttenoire:2023bqy,Ahmadvand:2023lpp,An:2023jxf,Wang:2023bbc}, domain walls\,\cite{Kitajima:2023cek,Guo:2023hyp,Blasi:2023sej,Gouttenoire:2023ftk,Barman:2023fad,Lu:2023mcz,Li:2023tdx,Du:2023qvj,Babichev:2023pbf,Gelmini:2023kvo,Zhang:2023nrs}, primordial fluctuations\,\cite{Franciolini:2023pbf,Vagnozzi:2023lwo,Franciolini:2023wjm,Inomata:2023zup,Cai:2023dls,Wang:2023ost,Ebadi:2023xhq,Gouttenoire:2023nzr,Liu:2023ymk,Abe:2023yrw,Unal:2023srk,Yi:2023mbm,Firouzjahi:2023lzg,Salvio:2023ynn,You:2023rmn,Bari:2023rcw,Ye:2023xyr,HosseiniMansoori:2023mqh,Cheung:2023ihl,Balaji:2023ehk,Jin:2023wri,Bousder:2023ida,Das:2023nmm,Zhao:2023joc,Ben-Dayan:2023lwd,Jiang:2023gfe,Liu:2023pau,Yi:2023tdk,Frosina:2023nxu,Bhaumik:2023wmw,Yuan:2023ofl,Gorji:2023sil},``audible" axions\,\cite{Figueroa:2023zhu,Geller:2023shn} and other more exotic scenarios\,\cite{Depta:2023qst,Yang:2023aak,Li:2023yaj,Lambiase:2023pxd,Borah:2023sbc,Datta:2023vbs,Murai:2023gkv,Niu:2023bsr,Choudhury:2023kam,Cannizzaro:2023mgc,Zhu:2023lbf,Aghaie:2023lan,He:2023ado}. If any such mechanism is in operation, one may expect there also to be an admixture of GWs from astrophysical SMBH binaries.~\footnote{See, for example,~\cite{Bian:2023dnv,Figueroa:2023zhu,Wu:2023hsa}}

In this paper, we perform a Multi-Model Analysis (MMA), comparing the qualities of fits to the NANOGrav 15-year (NG15) data invoking SMBH binaries, driven either by GWs alone or together with environmental effects, with each of the proposed fundamental physics sources. We also explore fits postulating combinations of SMBH binaries with each of the fundamental physics mechanisms. Our MMA adopts a common statistical approach to all the proposed sources based directly upon the NANOGrav probability density function (PDF) in each frequency bin. 
Furthermore, in order to help disentangle cosmological and astrophysical sources, we highlight some experimental features that could in the future be used to distinguish between them. These include fluctuations in the GW signal between different frequency bins (including the possible observation of individual GW sources) and the behavior of the SGWB spectrum at higher frequencies. For example, whereas SMBH models predict observable fluctuations that increase at higher frequencies, generic fundamental physics sources would not predict large bin-by-bin fluctuations, and some predict an extended spectrum that would be detectable at higher frequencies, whereas others predict a sharp fall-off above the range where PTAs are sensitive.

The structure of this paper is as follows. In Section~\ref{sec:SMBH} we introduce the astrophysical SMBH binary model, including the possibility of energy loss through environmental effects, and discuss the observability of bin-to-bin fluctuations and individual binary sources. Then, in Section~\ref{sec:cosmo} we introduce the different cosmological models we consider, based on fundamental physics scenarios, and present likelihood contours for each of the scenarios with and without a SMBH binary background, taking into account other relevant observational constraints. In Section~\ref{sec:comp} we compare the qualities of the fits within the various model scenarios, and we conclude in Section~\ref{sec:concl}. Some technical details are gathered in the appendix.

\section{Supermassive Black Holes}
\label{sec:SMBH}

The default interpretation of the SGWB postulates a population of tight astrophysical  SMBH binaries. We estimate the SMBH binary population as in~\cite{Ellis:2023dgf}: We use the extended Press-Schechter formalism~\cite{Press:1973iz,Bond:1990iw,Lacey:1993iv}, which depends on the rate $R_h$ of coalescences of galactic halos of masses $M_{1,2}$, the stellar mass-BH mass relation arising from observations of inactive galaxies~\cite{Kormendy:2013dxa,2015ApJ...813...82R} and the observed stellar mass-halo mass relation~\cite{Girelli:2020goz}. The merger rate of SMBHs of masses $m_{1,2}$ is
\bea
    \frac{\td R_{\rm BH}}{\td m_1 \td m_2} \approx & \,p_{\rm BH} \int \td M_1 \td M_2 \, \frac{\td R_h}{\td M_1 \td M_2} \\
    & \times p_{\rm occ}(m_1|M_1,z) p_{\rm occ}(m_2|M_2,z)  \,, 
\eea
where we incorporate a probability $p_{\rm BH}$ for the pair of SMBHs to approach each other sufficiently closely to merge~\cite{Ellis:2023owy} and $p_{\rm occ}$ relates the BH mass to the halo mass including the observed scatter~\cite{Ellis:2023dgf}. The interactions with their environments that allow the binaries to overcome the ``final parsec problem’’ are not fully understood~\cite{Begelman:1980vb}, and we regard $p_{\rm BH}$ as a phenomenological parameter to be determined using the PTA data. For simplicity, we assume $p_{\rm BH}$ to be constant.

\subsection{GW-driven binaries}

The spectrum of the GW signal from the SMBH population in general includes large fluctuations due to the small number of dominant and deviates from the naive power-law approximation that predicts~\cite{Phinney:2001di} the spectral index $\gamma = 13/3$. Following the analysis of Ref.~\cite{Ellis:2023dgf}, we compute the PDFs of the GW signal from SMBH binaries in the frequency bins used in the NG15 analysis and calculate their overlaps with the posterior PDFs (``violins") from the Hellings–Downs-correlated free spectrum analysis of the NG15 data~\cite{NANOGrav:2023gor}. Assuming that the evolution of the SMBH binaries is driven solely by energy loss due to GW emission, we find $p_{\rm BH} = 0.07_{-0.07}^{+0.05}$, which is consistent with an earlier analysis of IPTA data~\cite{Ellis:2023owy}, but the quality of the best fit is poor.

\begin{figure}
    \centering
    \includegraphics[width=0.95\columnwidth]{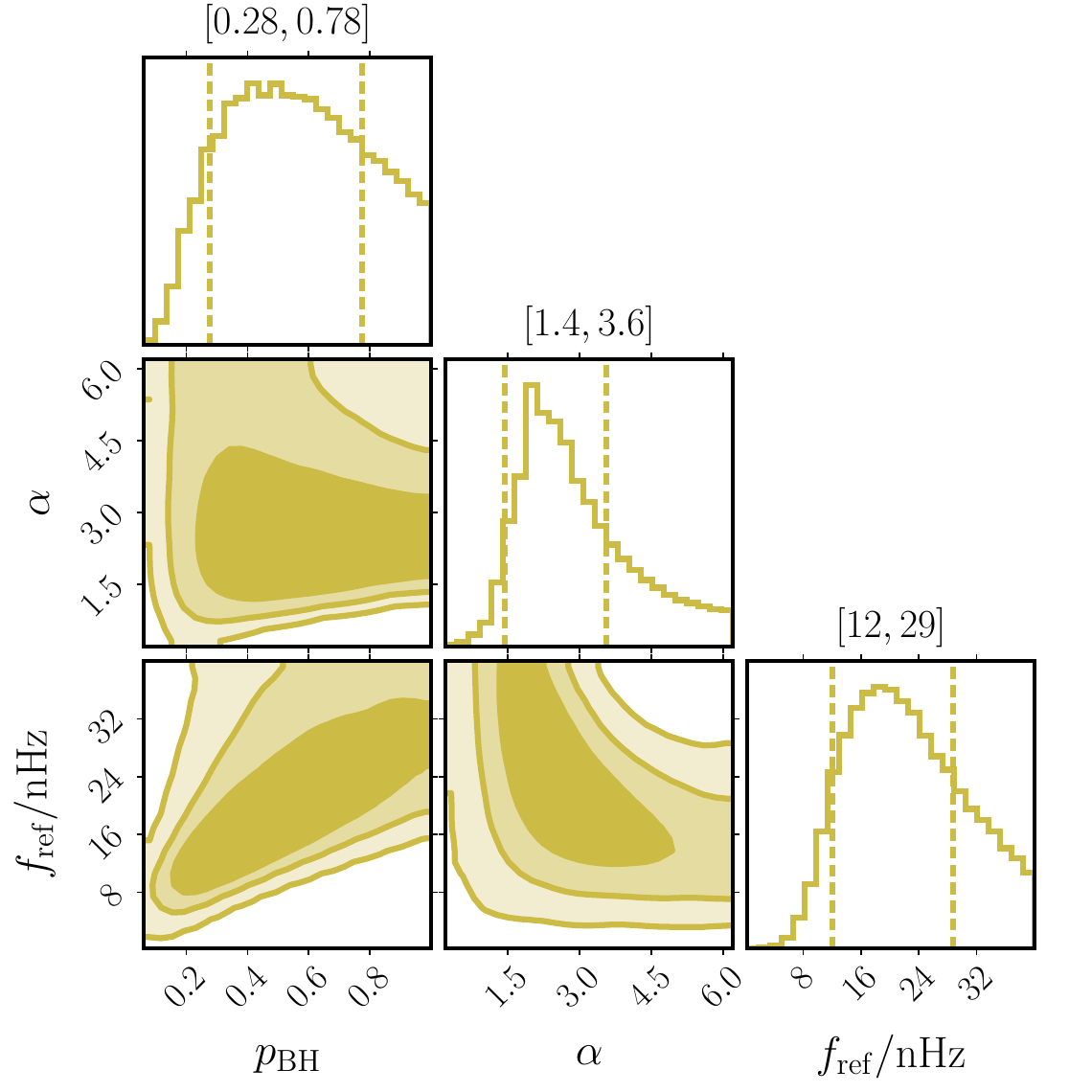}
    \caption{The posterior probability distribution of the fit of the SMBH binary model with environmental effects to the NG15 data. The contours enclose the $1\sigma$, $2\sigma$, and $3\sigma$ CL regions. On top of each column, we report $1\sigma$ CL ranges.}
    \label{fig:SMBH_posteriors}
\end{figure}

\subsection{Environmental effects}

It is natural to consider the possibility that environmental effects could affect the evolution of SMBH binaries while they are emitting GWs in the nHz frequency range. We parametrize the binary energy loss via environmental effects as~\cite{Ellis:2023dgf}
\be \label{eq:gas}
    \frac{t_{\rm env}}{t_{\rm GW}} = \left(\frac{f_r}{f_{\rm GW}}\right)^{\alpha} , \quad 
    f_{\rm GW} = f_{\rm ref} \left(\frac{\cM}{10^9 \Msun}\right)^{-\beta} , 
\ee
where $t_{\rm env}$ and $t_{\rm GW}$ are the timescales for energy loss via environmental effects and GWs, respectively, $f_{\rm ref}$ is a reference frequency, ${\cal M}$ is the binary chirp mass, and $\alpha$ and $\beta$ are phenomenological parameters. We take $\beta = 0.4$~\footnote{Our results are not sensitive to this choice: see~\cite{Ellis:2023dgf}.} and treat $p_{\rm BH}, \alpha$ and $f_{\rm ref}$ as parameters to be constrained using PTA data. This three-parameter model provides a significantly better fit to the NG15 data than the single-parameter GW-only model, with $-2 \Delta \ell = -11.3$ relative to the GW-driven SMBH model.\footnote{{We estimate the log-likelihood for observing the data $\vec{d}$ given a model characterised by $\vec{\theta}$ by
\be
    \ell(\vec{d}|\vec{\theta}) = \sum_j \ln \int \td \Omega P_{{\rm ex},j}(\Omega|\vec{d}) P_{{\rm th},j}(\Omega|\vec{\theta}) \,,
\ee
where the sum runs over the first 14 NG15 bins and $P_{{\rm ex},j}$ and $P_{{\rm th},j}$ denote the probability distributions of $\Omega$ in bin $j$ in the NG15 data and in the given model.}} The posterior probabilities for the three parameters, computed in Ref.~\cite{Ellis:2023dgf}, are shown in Fig.~\ref{fig:SMBH_posteriors}. The best fit is at 
\be
    p_{\rm BH} = 0.84,\quad \alpha = 2.0, \quad f_{\rm ref} = 34\,{\rm nHz} \,.
\ee
Throughout the rest of the paper, this best-fit environmentally driven SMBH scenario sets the baseline to which other models are compared.

We note that our analysis assumes circular orbits that emit GWs at twice the orbital frequency. The eccentricity of the orbits would affect the GW spectrum by introducing higher harmonics, increasing the total power emitted in GWs and modifying the frequency spectral index~\cite{Enoki:2006kj}.\footnote{There is no indication of eccentric binaries in current (O3) LIGO/Virgo/KAGRA (LVK) data~\cite{LIGOScientific:2023lpe}.}  However, big eccentricities $e>0.9$ would lead to an attenuation of the background due to the acceleration of the binary inspiral~\cite{Kelley:2017lek}. The GW signal from SMBHs could also be affected by modifications to small-scale structures compared to standard $\Lambda$CDM, which could lead to the earlier formation of galaxies and SMBHs. The observation of high-redshift galaxies with surprisingly high stellar masses by the JWST~\cite{2022arXiv220802794N, 2022ApJ...940L..55F, 2023Natur.616..266L, 2023ApJS..265....5H,2023arXiv230406658H} provides a potential hint for this possibility~\cite{Liu:2022bvr, Menci:2022wia, Biagetti:2022ode, Hutsi:2022fzw, Parashari:2023cui, Hassan:2023asd, Gouttenoire:2023nzr, Guo:2023hyp}. 

\begin{figure}
    \centering
    \includegraphics[width=0.75\columnwidth]{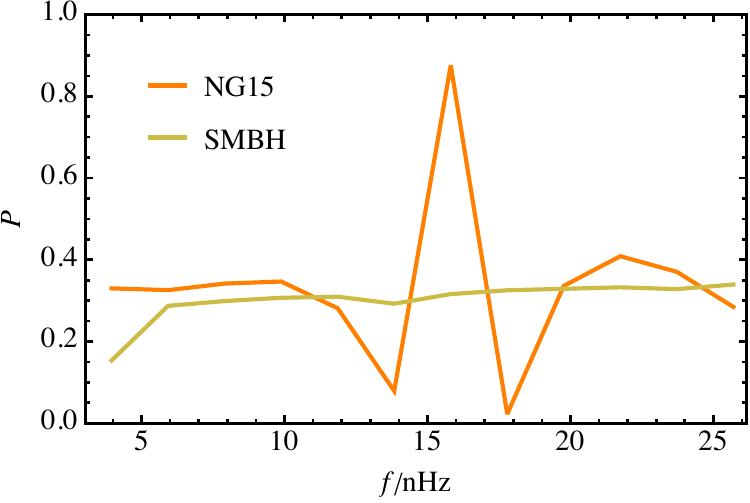}
    \caption{The probability~\eqref{eq:Pfluct} for upward fluctuations in the SMBH model with environmental effects (green line) and the fluctuations measured in the NG15 data (orange line).}
    \label{fig:Pfluct}
\end{figure}

\subsection{Fluctuations in the GW spectrum}

As already mentioned, the SMBH signal may exhibit significant fluctuations between frequency bins, which appear because around half of the signal strain is produced by ${\cal O}(10)$ sources. {Since the spectral fluctuations in the SGWB for every cosmological BSM model are expected to be negligible, the appearance of fluctuations would be a smoking gun for an astrophysical component in the signal.}

Spectral fluctuations can be quantified by the second finite difference
\bea\label{eq:Delta2_Omega}
    \Delta^2\Omega (f) 
    &\equiv \Omega(f+\delta f) + \Omega(f-\delta f) - 2\Omega(f) \, ,
\eea
where the step size $\delta f$ can, for instance, be taken to correspond to the width of a bin. The probability of an upward fluctuation in the $i$th bin can be estimated as~\footnote{Neglecting correlations between bins, this probability is given by
\bea
    P(-\Delta^2\Omega(f_i) > \Omega_{\rm th} )
    =  & \int \td \Omega \td \Omega' P_{i+1}(\Omega-\Omega') P_{i-1}(\Omega') \\ 
    &\times \int_{2\Omega_i - \Omega > \Omega_{\rm th}(\Omega,f_i)} \td \Omega_i P_i(\Omega_i)
\eea
where $P_{i}$ denote the theoretical $\Omega_{\rm GW}$ distributions in individual bins (see~\cite{Ellis:2023owy,Ellis:2023dgf}).}
\be \label{eq:Pfluct}
    P_{\rm fluct, i} \equiv P(-\Delta^2\Omega(f_i) > \Omega_{\rm th}(f_i) ) \, ,
\ee
where $f_i$ is the bin frequency and we choose
\be
    \Omega_{\rm th}(f_i) = \iota \left[\Omega (f_{i-1}) + \Omega (f_{i+1})\right] \,,
\ee
with $\iota = 0.2$. This choice of threshold can help distinguish SMBH binaries from cosmological GW sources, which are expected to produce smooth spectra. We find that for all of the cosmological sources discussed in the next Section the probability~\eqref{eq:Pfluct} is zero in all NG15 bins. When the GW signal arises from an admixture of cosmological sources and SMBH binaries the probability of fluctuations will be smaller than in the pure SMBH model. The optimal threshold $\Omega_{\rm th}$ that can filter out cosmological models depends on the frequency binning -- while for smooth cosmological signals, the finite difference \eqref{eq:Delta2_Omega} decreases with increased spectral resolution (since $\Delta^2\Omega(f) \propto \delta f^2$ as $\delta f \to 0$), it can be sizeable for SMBH binaries as long as their frequency changes less than $\delta f$ during the observational period.
 
In Fig.~\ref{fig:Pfluct} we show the probability~\eqref{eq:Pfluct} for the best-fit parameters of the SMBH binary model with environmental effects. We find that $P\approx 0.3$ across all the bins and the probability of not having fluctuations that exceed this threshold in any of the 11 bins considered in Fig.~\ref{fig:Pfluct} is roughly $\prod P_{\rm fluct, i} \approx 1.3\%$, while the expected number of upward fluctuations in the NG15 bins is $\sum_i  P_{\rm fluct, i} \approx 4$ for both the NG15 data and the SMBH binary model. The NG15 data indicates that the 8th bin at $f\approx 16\,$nHz shows an upward fluctuation at the 91\% CL.

\begin{figure}
\centering
\includegraphics[width=0.8\columnwidth]{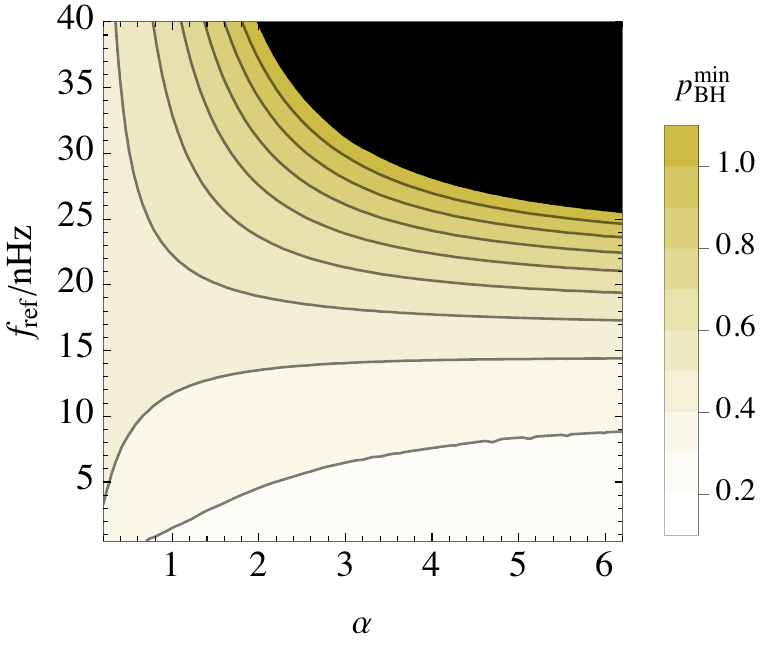}
\caption{Contours of the minimal merger efficiency $p_{\rm BH}^{\rm min}$ required for the probability of finding the candidate event at $f \sim 4 \rm{nHz}$ to exceed 5\% for different values of the parameters of the environmental effects. In the black region, the probability of the candidate event is less than 5\% for $p_{\rm BH}\leq 1$.}
\label{fig:Pmin}
\end{figure}

\subsection{Detecting individual binaries}

A nearby SMBH binary will stand out from the background as a resolvable binary. Intriguingly, the model predicts that the most probable frequency for an individual binary to be detectable is $\sim 4$~nHz~\cite{Ellis:2023dgf}, where there is evidence for an upward fluctuation in both NANOGrav and EPTA data~\cite{NANOGrav:2023pdq,Antoniadis:2023bjw}. If the SMBH binaries are driven by GW emission only the probability of observing such an event in the present NANOGrav data is less than$\sim 5\%$, but such an event is easier to accommodate when environmental energy loss is included in the SMBH binary model, where the probability is $\sim 23\%$. In Fig.~\ref{fig:Pmin} we show contours of the minimal merger efficiency, $p_{\rm BH}^{\rm min}$, for which the probability of finding such an event is larger than 5\% for different values of the environmental parameters $\alpha$ and $f_{\rm ref}$.~\footnote{In addition to bin-to-bin fluctuations and resolvable binaries, other characteristic features of the SMBH models include the possibilities of observable circular polarization~\cite{Sato-Polito:2021efu,Ellis:2023owy} and anisotropies~\cite{Sato-Polito:2023spo}.}

\section{Cosmological sources}
\label{sec:cosmo}

In this Section, we discuss SGWB predictions in various cosmological scenarios invoking particle physics beyond the Standard Model (BSM). In these scenarios, the GWs are produced in the early Universe and their present abundance is
\bea\label{eq:OmegaGWtoday}
    &\Omega_{\rm GW} h^2 = \left[\frac{a(T)}{a_0}\right]^4 \left[\frac{H(T)}{H_0/h}\right]^2 \Omega_{\rm GW}(T) \\ 
    &\approx 1.6\times 10^{-5} \left[ \frac{g_*(T)}{100} \right] \!\left[ \frac{g_{*s}(T)}{100} \right]^{\!-\frac43} \Omega_{\rm GW}(T) \,,
\eea
where $T$ denotes the temperature at which the GWs were produced and $\Omega_{\rm GW}(T)$ their abundance at that moment, $a$ is the scale factor and $H$ the Hubble rate, the subscript $0$ refers to their present day values, and $g_*$ and $g_{*s}$ are the effective numbers of relativistic energy and entropy degrees of freedom. The present frequency corresponding to the scale entering the Hubble horizon at temperature $T$ is
\bea\label{eq.fre}
    f_H(T)\!&=\frac{a(T)}{a_0} \frac{H(T)}{2\pi} \\
    &\approx\!2.6\!\times\!10^{-8}\,{\rm Hz} \left[ \frac{g_*(T)}{100} \right]^{\frac12}\!\left[ \frac{g_{*s}(T)}{100} \right]^{\!-\frac13}\!\frac{T}{\rm GeV} \,.
\eea
For short-lasting, so-called causality-limited GW sources, the GW spectrum at $f\ll f_H$ scales as $\Omega_{\rm GW}(f\ll f_H)= \Omega_{\rm CT}(f) \propto f^3$, assuming radiation domination~\cite{Caprini:2009fx,Domenech:2020kqm}. This behavior is affected by deviations from pure radiation domination, e.g., around the QCD phase transition. We use the results tabulated in Ref.~\cite{Franciolini:2023wjm} for the causality tail, $\Omega_{\rm CT}(f)$.

\subsection{Cosmic (super)strings}

Cosmic strings are one-dimensional topological defects that are predicted in many BSM scenarios~\cite{Hindmarsh:1994re,Jeannerot:2003qv,King:2020hyd}. Once produced, strings reach a scaling solution where their density is a small fraction of the total cosmological density while the excess energy goes into production of closed loops that then decay into GWs. This process continues throughout the evolution of the Universe and results in a very broad and relatively flat spectrum.
We focus here on (super)strings that may arise from string theory~\cite{Dvali:2003zj,Copeland:2003bj} or during  confinement in pure Yang-Mills theories~\cite{Yamada:2022aax,Yamada:2022imq}. Their main phenomenological difference from the standard case is a lower intercommutation probability $p$, which diminishes the loop production and increases the resulting string density and GW signal amplitude by a factor $p^{-1}$~\cite{Sakellariadou:2004wq,Blanco-Pillado:2017rnf}.  

Our calculation of the spectrum closely follows~\cite{Ellis:2023tsl}. We start with the velocity-dependent one-scale model~\cite{Martins:1995tg, Martins:1996jp, Martins:2000cs, Avelino:2012qy, Sousa:2013aaa}, which we use to compute the correlation length of the network $L$ and mean string velocity $\bar{v}$ by solving the equations
\bea \label{eq:vos}
    \frac{dL}{dt} &= (1+\bar{v}^2)\,HL + \frac{\tilde{c}\bar{v}}{2} \,, \\
    \frac{d \bar{v}}{dt} &= (1-\bar{v}^2)
    \left[\frac{k(\bar{v})}{L} - 2H\,\bar{v}\right] \,,
\eea
where
\be
    k(\bar{v}) = \frac{2\sqrt{2}}{\pi}(1-\bar{v}^2)(1+2\sqrt{2}\bar{v}^3)
\left(\frac{1-8\bar{v}^6}{1+8\bar{v}^6}\right) \ ,
\ee
and $\tilde{c} \simeq 0.23$ is the loop chopping efficiency~\cite{Martins:2000cs}.

Following recent numerical simulations~\cite{Blanco-Pillado:2011egf, Blanco-Pillado:2013qja, Blanco-Pillado:2015ana, Blanco-Pillado:2017oxo, Blanco-Pillado:2019vcs, Blanco-Pillado:2019tbi} we focus on large loops as the main GW sources. We include a prefactor ${\cal F} \sim 0.1$ to describe the fraction of energy in these loops and a factor $f_r=\sqrt{2}$ accounting for the energy going to peculiar velocities of loops~\cite{Auclair:2019wcv} and not contributing to radiation.

Loops cut from the network oscillate losing energy to GWs which reduces their length over time:
\be \label{eq:lloop}
    l (t) = \alpha_L L \left(t_i \right) - \Gamma G \mu \left(t -t_i \right) \ .
\ee
Where we use $\alpha_L \approx 0.37$, the total emission power $\Gamma \approx 50$~\cite{Burden:1985md, Blanco-Pillado:2017oxo, Blanco-Pillado:2017rnf} and $t_i$ is the time at which the given loop was created.

\begin{figure}
\centering
\includegraphics[width=0.95\columnwidth]{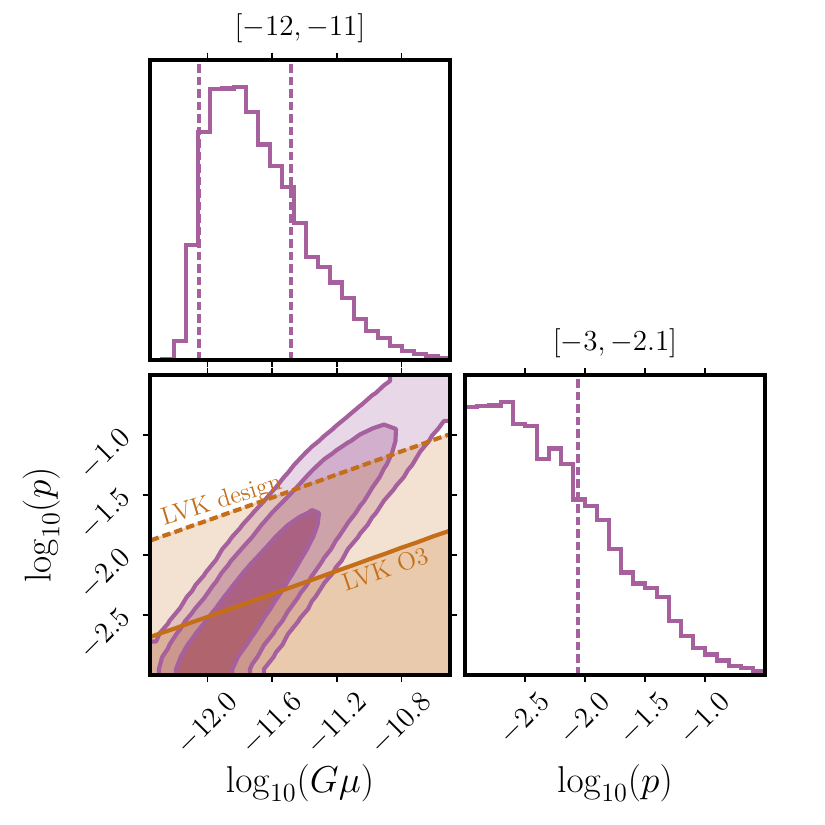}
\caption{The posterior probability distribution of the cosmic (super)string model fit to the NG15 data. The current (O3) LVK constraint and prospective future LVK sensitivity are shown by the solid and dashed lines.}
\label{fig:CSfit}
\end{figure}

Calculation of the spectrum involves the integration of emissions from all contributing loops:
\be
\label{eq:GWdensity}
    \Omega_{\rm GW}^{(1)}(f) = \frac{16\pi}{3 f} \frac{\mathcal{F}}{f_r} \frac{G\mu^2 }{H_0^2} \frac{ \Gamma}{\zeta \alpha_L} \int_{t_F}^{t_0}\!d\tilde{t}\; n(l,\tilde{t}) \, , 
\ee
where
\be
    n(l,\tilde{t}) = \frac{\Theta(t_i - t_F)}{\alpha_L \dot{L}(t_i)+\Gamma G\mu} \frac{\tilde{c} v(t_i)}{L(t_i)^4}  \bigg[\frac{a(\tilde{t})}{a(t_0)}\bigg]^5 \bigg[\frac{a(t_i)}{a(\tilde{t})}\bigg]^3 \, .
\ee
The formation time of the loops has to be found solving Eq.~\eqref{eq:lloop} and restricting to loops emitting at frequency $f$ whose length is $l(\tilde{t},f)=\frac{2}{f}\frac{a(\tilde{t})}{a(t_0)}$.
The Heaviside function assures that we integrate from only the time $t_F$ when the network first reaches scaling after its production. We ensure that the total emitted power equals $\Gamma$ with the function $\zeta=\sum k^{-4/3}$, where we assume that the GW emission is dominated by cusps.
Finally, we sum over the emission modes using
\be
    \Omega_{\rm GW}^{(k)}(f) =k^{-q}\,\Omega_{\rm GW}^{(1)}(f/k) \, ,
\ee
and the prescription from ref. ~\cite{Cui:2019kkd} that allows us to sum over a very large number of modes, which is necessary to compute the spectrum accurately~\cite{Cui:2019kkd, Blasi:2020wpy, Gouttenoire:2019kij}.

\begin{figure}
\centering
\includegraphics[width=0.95\columnwidth]{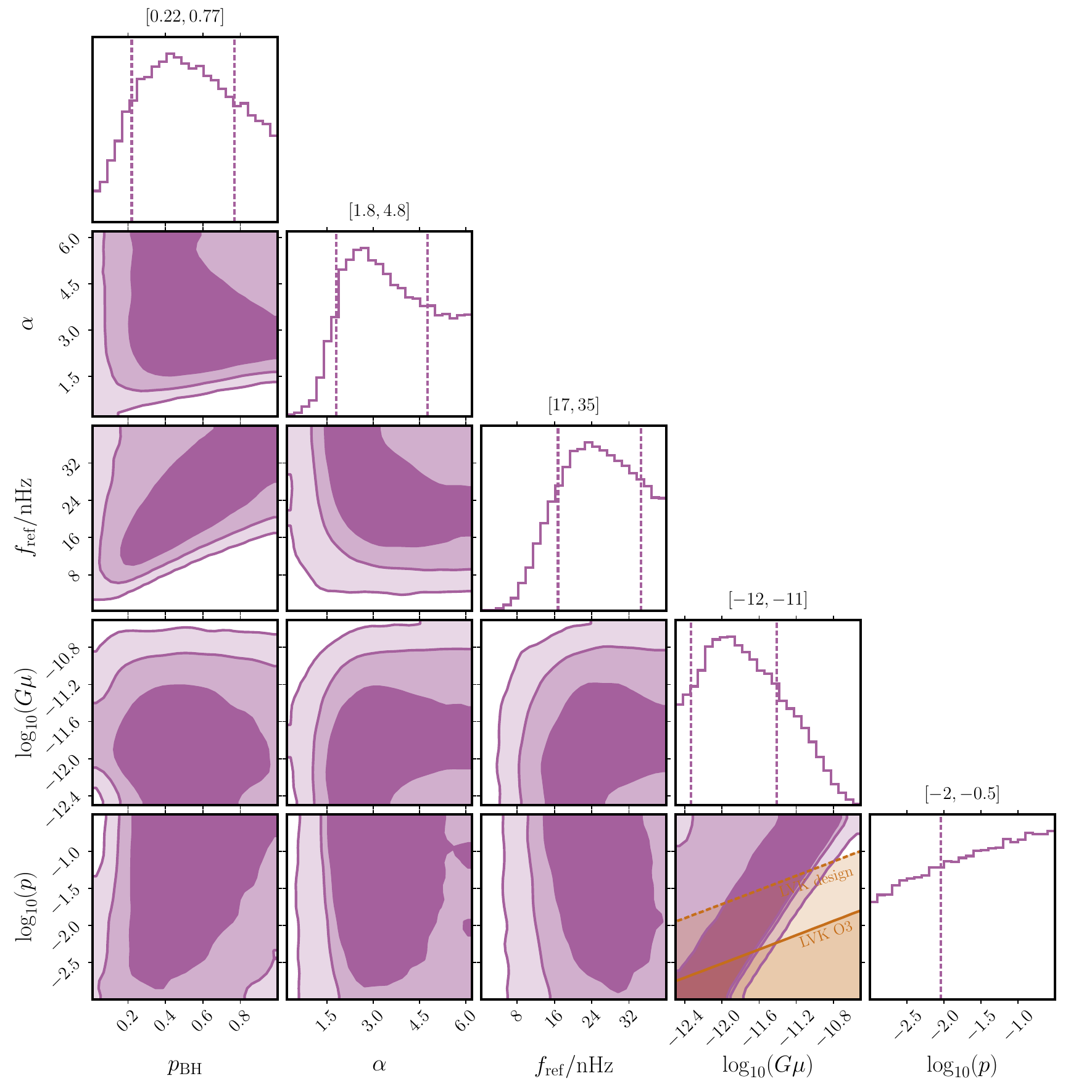}
\caption{As in Fig.~\ref{fig:CSfit}, for a (super)string fit to NG15 data including also SMBH binaries with environmental effects.}
\label{fig:CSwSMBHfit}
\end{figure}

In fig.~\ref{fig:CSfit} we show the posterior distributions for the string tension, $G\mu$, and the intercommutation probability, $p$, together with the constraints from current (O3) LIGO/Virgo/KAGRA (LVK) data~\cite{KAGRA:2021kbb} and the design sensitivity of future LVK data~\cite{LIGOScientific:2014pky,LIGOScientific:2016fpe,LIGOScientific:2019vic}, assuming that the SMBH background is negligible. The best pure (super)string fit not in conflict with current LVK data has
\be
    G\mu = 2\times 10^{-12} \,, \quad p = 6.3\times 10^{-3} \,, 
\ee
with $-2 \Delta \ell=1.5$ compared to the SMBH baseline model including environmental effects. Fig.~\ref{fig:CSwSMBHfit} shows results from a combined fit including the possibility of a SMBH background.
The best fit, in this case, is at
\bea
    &G\mu = 1.2\times 10^{-12} \,, \quad p = 3.3\times 10^{-3} \,, \\
    &p_{\rm BH}=0.42 \,, \quad f_{\rm ref}=24 \, {\rm nHz}\,, \quad \alpha=4.4 \,,  
\eea
with $-2 \Delta \ell=-0.7$. Each model individually can give a comparable fit, and indeed we find that a combined fit with significant contributions from both gives a noticeably better fit, although the improvement is not very significant statistically. A significant contribution from SMBH binaries also results in a non-negligible probability of finding a candidate event at $f \sim 4$~nHz, which in this case is 5\%.

\subsection{Phase transitions}

Cosmological first-order phase transitions generate GWs in bubble collisions and by motions of inhomogeneities in the fluid~\cite{Witten:1984rs,Hogan:1986qda}. Given the large amplitude of the NANOGrav signal, we focus on very strong transitions that accommodate this feature naturally. We assume that the energy of the transition dominates over the background, so that its strength  $\alpha \gg 1$, in which case its value does not play an important role in the calculation of the signal. It was recently shown in~\cite{Lewicki:2022pdb} that in such a case the energy budget of the transition does not play an important role, as fluid-related sources~\cite{Kamionkowski:1993fg,Hindmarsh:2015qta,Hindmarsh:2016lnk,Hindmarsh:2017gnf,Ellis:2018mja,Cutting:2019zws,Hindmarsh:2019phv,Ellis:2020awk,Nakai:2020oit} and the contributions from bubble collisions~\cite{Kosowsky:1992vn,Cutting:2018tjt,Ellis:2019oqb,Lewicki:2019gmv,Cutting:2020nla,Lewicki:2020jiv,Giese:2020znk,Ellis:2020nnr,Lewicki:2020azd,Wang:2022wwj} behave very similarly.

\begin{figure}
\centering
\includegraphics[width=0.95\columnwidth]{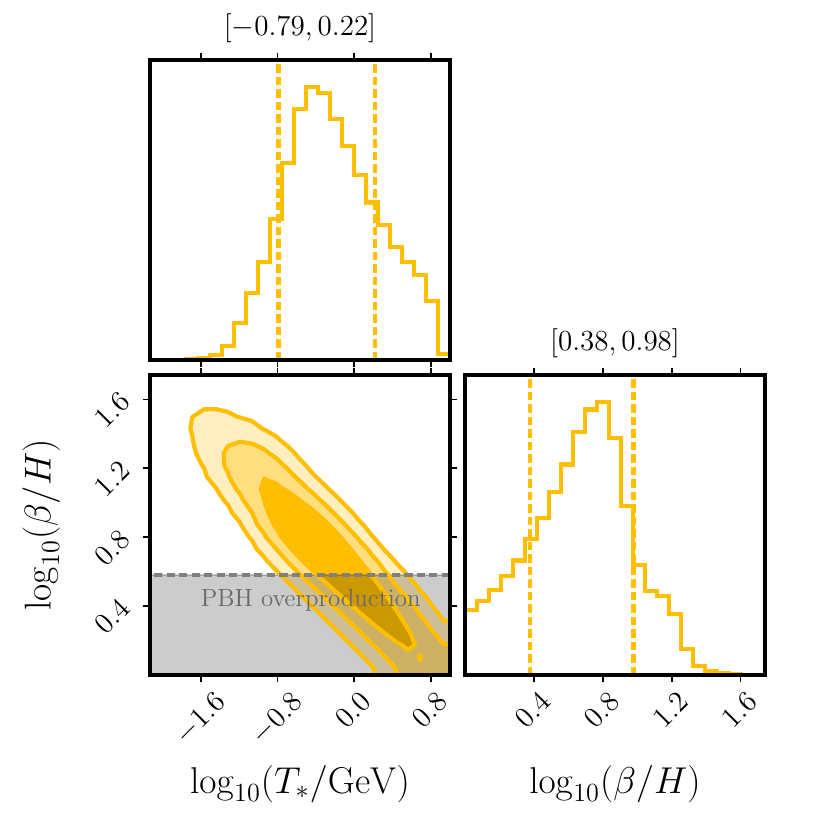}
\caption{The posterior probability distribution of the phase transition fit to the NG15 data. The constraint from the production of PBHs is shown in grey.}
\label{fig:PTfit}
\end{figure}

The resulting GW spectrum is a broken power-law that may be parametrized as 
\be \label{eq:Omega_PT}
    \Omega_{\textrm{GW}}(f,T_*) = \left[\frac{\beta}{H}\right]^2 \frac{A (a+b)^c S_H(f,f_H(T_*))}{\left(b \left[\frac{f}{f_p}\right]^{\!-\frac{a}{c}} + a \left[\frac{f}{f_p}\right]^{\frac{b}{c}}\right)^{\!c}} \,,
\ee
where $\beta$ is the timescale of the transition, $T_*$ is the temperature reached after the transition,
$a=b=2.4$, $c=4$, $A=5.1\times 10^{-2}$ and the peak frequency $f_p = 0.7 f_H(T_*) \beta/H$~\cite{Lewicki:2022pdb}. The function
\be \label{eq:SH}
    S_H(f,f_H) = \left(1+\left[ \frac{\Omega_{\rm CT}(f)}{\Omega_{\rm CT}(f_H)}\right]^{-\frac{1}{\delta}}\left[\frac{f}{f_H}\right]^{\frac{a}{\delta}}\right)^{\!-\delta}
\ee
models the transition of the spectrum to the causality tail at scales larger than the horizon at the transition time~\cite{Caprini:2009fx}. The parameter $\delta$ determines how quickly the spectrum evolves towards the causality tail and could be determined by a simulation of the strong phase transition taking expansion into account. We fix $\delta = 1$, having verified its value does not have a large impact on the results. 

\begin{figure}
\centering
\includegraphics[width=0.95\columnwidth]{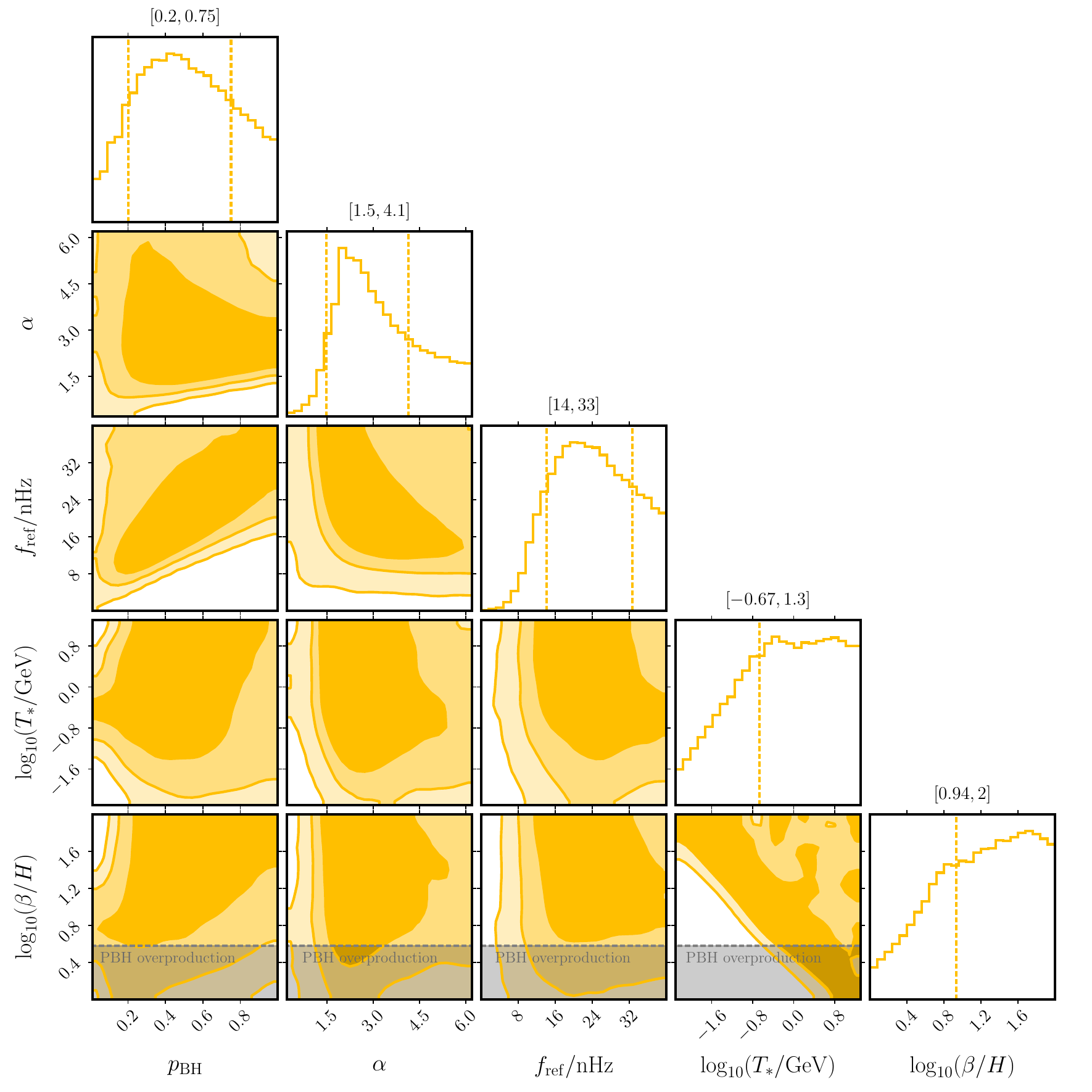}
\caption{As in Fig.~\ref{fig:PTfit} for a phase transition fit to NG15 data including also SMBH binaries with environmental effects.}
\label{fig:PTwSMBHfit}
\end{figure}

In Fig.~\ref{fig:PTfit} we show the posteriors of phase transition fit to the NG15 data while Fig.~\ref{fig:PTwSMBHfit} includes also the background from SMBH binaries. The best fit to a pure phase transition signal is for
\be \label{eq:PTbestfit}
    T_* = 0.34 \, {\rm GeV}\,, \quad \beta/H = 6.0\,, 
\ee
resulting in $-2 \Delta \ell =-2.3$ compared to the SMBH baseline model. Including SMBHs with environmental effects, we find that the best fit is dominated by the phase transition signal with $p_{\rm BH} \approx 0$. Searching for points capable of providing the candidate event at $4$\,nHz by requiring its probability to be $\geq 5\%$ we find the best-fit point
\bea
    &T_*=2.6 \,{\rm GeV}\,, \quad \beta/H =4.0 \\
    &p_{\rm BH}=0.28\, , \; f_{\rm ref}=20 \, {\rm nHz}\, , \; \alpha=2.6\, ,  
\eea
which gives $-2 \Delta \ell =0.2$. 

Strong first-order phase transitions can also lead to the formation of primordial black holes (PBHs) if they are sufficiently slow~\cite{Hawking:1982ga, Kodama:1982sf, Liu:2021svg, Lewicki:2023ioy, Gouttenoire:2023naa}. In Figs.~\ref{fig:PTfit} and \ref{fig:PTwSMBHfit} we show the constraint $\beta/H \gtrsim 4$ from overproduction of PBHs~\cite{Lewicki:2023ioy}.\footnote{A slightly stronger bound, $\beta/H \gtrsim 6$, was found in~\cite{Gouttenoire:2023naa}.} We see that most of the 1$\sigma$ region of the fit including only the phase transition signal is allowed by this constraint. However, as pointed out in~\cite{Gouttenoire:2023bqy}, it is interesting that the NG15 fit is compatible with a large PBH abundance. These PBHs would be around the stellar mass range, where their abundance is constrained by optical lensing~\cite{EROS-2:2006ryy, Macho:2000nvd, Zumalacarregui:2017qqd,Gorton:2022fyb,Petac:2022rio,DeLuca:2022uvz}, GW observations~\cite{Raidal:2017mfl, Ali-Haimoud:2017rtz, Raidal:2018bbj, Vaskonen:2019jpv,  Hutsi:2020sol, Franciolini:2022tfm} and accretion~\cite{Ricotti:2007au, Horowitz:2016lib, Ali-Haimoud:2016mbv, Poulin:2017bwe, Hektor:2018qqw, Hutsi:2019hlw, Serpico:2020ehh} (for a review see e.g.~\cite{Carr:2020gox}). Nevertheless, a subdominant PBH abundance, $\mathcal{O}(0.1)$\% of DM, might be related to the BH binary mergers detected by LVK~\cite{Vaskonen:2019jpv,DeLuca:2020jug,Raidal:2017mfl,Raidal:2018bbj,Ali-Haimoud:2017rtz,DeLuca:2020qqa,Hutsi:2020sol,Franciolini:2021tla,Clesse:2020ghq,Franciolini:2022tfm} and can be further probed with future GW observatories~\cite{DeLuca:2021hde, Pujolas:2021yaw, Urrutia:2023mtk, Franciolini:2023opt}.

\subsection{Domain walls}

\begin{figure}
    \includegraphics[width=0.95\columnwidth]{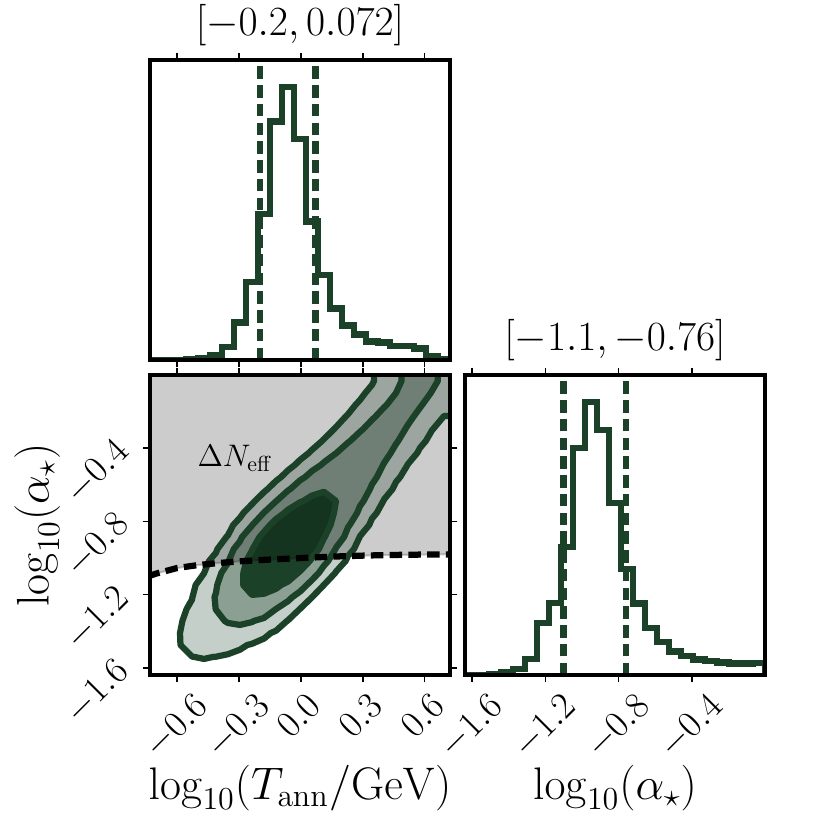}
    \caption{The posterior probability distribution of the DW fit to the NG15 data. The dashed line indicates the $\Delta N_{\rm eff}$ bound, which constrains the possibility of DWs annihilating completely into dark radiation.}
    \label{fig:new_DW}
\end{figure}

Domain walls (DWs)~\cite{Vilenkin:1984ib} are topological defects produced when a discrete symmetry in some BSM scenario is broken after inflation. During the scaling regime when the DW network expands together with its surroundings, the energy density is $\rho_{\rm DW}=c\, \sigma H$~\cite{Hiramatsu:2013qaa, Kibble:1976sj}, where $\sigma$ is the tension of the wall and $c = \mathcal{O}(1)$ is a scaling parameter. DWs emit GWs until they annihilate at a temperature $T = T_{\rm ann}$~\cite{Gelmini:1988sf,Coulson:1995nv,Larsson:1996sp,Preskill:1991kd}. The peak frequency of the resulting GW spectrum is given by the horizon size at the time of DW annihilation, $f_p = f_H(T_{\rm ann})$, and at frequencies $f\gg f_p$ the spectrum scales as $f^{-1}$. We approximate the GW spectrum at the formation time $T=T_{\rm ann}$ as~\cite{Hiramatsu:2013qaa,Ferreira:2022zzo}
\be
    \Omega_{\rm DW}(f,T_{\rm ann})
    = \frac{3 \epsilon \alpha_*^2}{8\pi} \!\left( \frac14\! \left[ \frac{\Omega_{\rm CT}(f_p)}{\Omega_{\rm CT}(f)}\right]^{\frac{1}{\delta}} \!\!+ \frac34\! \left[\frac{f}{f_p}\right]^{\!\frac{1}{\delta}}\right)^{\!\!-\delta} ,
\ee
where $\epsilon = \mathcal{O}(1)$ is an efficiency parameter and $\alpha_* \equiv \rho_{\rm DW}(T_{\rm ann})/\rho_r(T_{\rm ann})$ is the energy density in the domain walls relative to the radiation energy density $\rho_r$ at the annihilation moment. Following the results of numerical simulations~\cite{Hiramatsu:2013qaa}, we fix $\epsilon=0.7$ and $\delta = 1$.

\begin{figure}
    \centering
    \includegraphics[width=0.95\columnwidth]{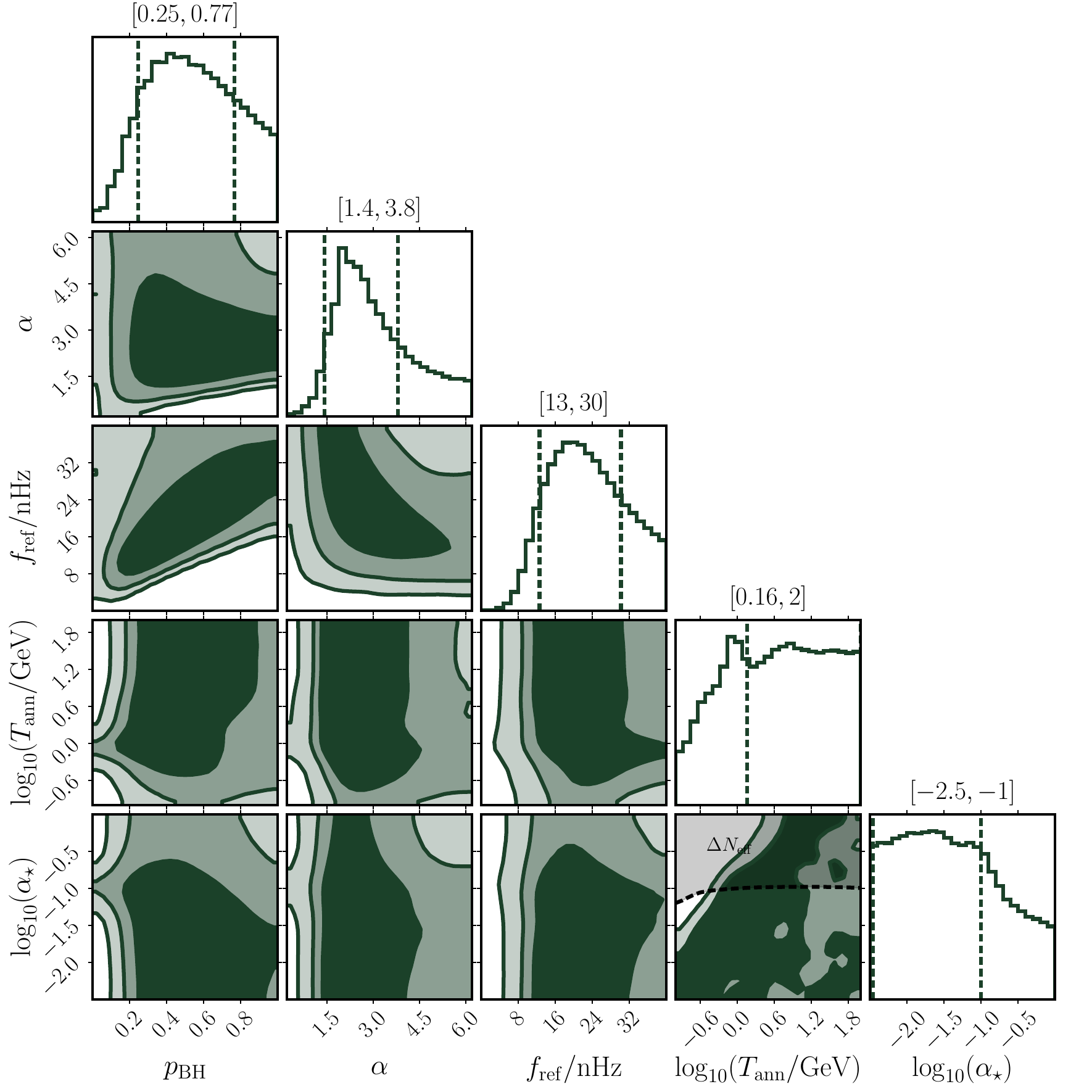}
    \caption{As in Fig.~\ref{fig:new_DW} for a domain wall fit to the NG15 data including also SMBH binaries with environmental effects.}
    \label{fig:DW_posteriors}
\end{figure}

The parameters of the DW model are the relative energy density in DWs, $\alpha_*$, and the temperature at which they annihilate, $T_{\rm ann}$. Since the DWs can constitute a big fraction of the total energy density, it is necessary to check that their annihilation respects the constraints imposed by Big Bang Nucleosynthesis (BBN) and the Cosmic Microwave Background (CMB).  If the DWs annihilate completely into dark radiation, the energy density can be expressed as the equivalent number of neutrino species~\cite{Ferreira:2022zzo}: $\Delta N_{\rm eff} = \rho_{\rm DW}(T_{\rm ann})/\rho_{\rm \nu}(T_{\rm ann}) = 13.6 \, g_{*}(T_{\rm ann})^{-\frac{1}{3}}\, \alpha_*$, which is constrained by BBN ($\Delta N_{\rm eff} < 0.33$)~\cite{Fields:2019pfx} and CMB ($\Delta N_{\rm eff} < 0.3$)~\cite{Planck:2018vyg,Ramberg:2022irf}. On the other hand, if the DWs annihilating into SM particles, BBN imposes $T_{\rm ann} > (4-5) \, {\rm MeV}$~\cite{Hasegawa:2019jsa} and CMB $T_{\rm ann} > 4.7 \, {\rm MeV}$~\cite{deSalas:2015glj} (see~\cite{Allahverdi:2020bys} for a review).

We have performed scans over the parameters of the DW model to fit the NG15 data. The posterior probability distributions for the DW model without a SMBH contribution are shown in Fig.~\ref{fig:new_DW}, and the best fit is at
\be
    T_{\rm ann} = 0.85\,{\rm GeV}\,, \quad \alpha_* = 0.11 \,, 
\ee
and has $-2 \Delta \ell=-3.1$ compared to the SMBH baseline model. These parameter values are allowed both if the DWs annihilate into SM particles. If they instead annihilate into dark radiation, the best fit is in tension with $\Delta N_{\rm eff}$ bound. However, the $1\sigma$ region of the posterior distribution extends to smaller values of $\alpha_*$ that are allowed by the $\Delta N_{\rm eff}$ bound.

Fig.~\ref{fig:DW_posteriors} shows a full set of posterior parameter distribution functions and two-parameter correlations for the fit to a combination of DWs and SMBH binaries. The best fit of the combined model is dominated by the DW contribution with $p_{\rm BH} \approx 0$. Imposing a lower bound on $p_{\rm BH}$ from Fig.~\ref{fig:Pmin}, the best fit is 
\bea \label{FWSMBHfit}
    &T_{\rm ann} = 1.12 \, {\rm GeV}\,, \quad \alpha_* = 0.074\,, \\
    &p_{\rm BH}=0.51\, , \; f_{\rm ref} = 22 \, {\rm nHz}\, , \; \alpha = 2.6\, ,
\eea
with $-2 \Delta \ell=-0.5$. 

Similarly to first-order phase transitions, the annihilation of the DW network can lead to the copious formation of PBHs~\cite{Ferrer:2018uiu,Gelmini:2023ngs}. The PBH formation may be so efficient that it excludes the parameter region preferred by the NG15 data~\cite{Gouttenoire:2023ftk}. However, further studies are needed to confidently estimate the PBH formation from DW networks.

\subsection{Scalar-induced GWs}
\label{sec:SIGWth}

Primordial scalar curvature fluctuations can source a SGWB at the second order of perturbation theory~\cite{Tomita:1975kj,Matarrese:1993zf,Acquaviva:2002ud,Mollerach:2003nq,Ananda:2006af,Baumann:2007zm,Domenech:2021ztg,Yuan:2021qgz}. A scalar-induced GW (SIGW) background that can be probed by PTAs requires sizable curvature perturbations that enter the horizon around the QCD phase transition. This scenario has received a lot of attention~\cite{Vaskonen:2020lbd,Chen:2019xse,DeLuca:2020agl,Bhaumik:2020dor,Inomata:2020xad,Kohri:2020qqd,Domenech:2020ers,Vagnozzi:2020gtf,Namba:2020kij,Sugiyama:2020roc,Zhou:2020kkf,Lin:2021vwc,Rezazadeh:2021clf,Kawasaki:2021ycf,Ahmed:2021ucx,Yi:2022ymw,Yi:2022anu,Dandoy:2023jot,Zhao:2023xnh,Ferrante:2023bgz,Cai:2023uhc} since the initial observation of a stochastic common-spectrum process by NANOGrav~\cite{NANOGrav:2020bcs} and several studies have recently appeared that update the available parameter space in the light of the most recent PTA data releases~\cite{Franciolini:2023pbf, Franciolini:2023wjm, Inomata:2023zup, Ebadi:2023xhq, Figueroa:2023zhu, Yi:2023mbm, Firouzjahi:2023lzg, You:2023rmn,Balaji:2023ehk,Zhao:2023joc,Yi:2023tdk}. The most stringent constraint on a SIGW scenario for PTAs arises from the overproduction of PBHs associated with large peaks in the primordial curvature power spectrum~\cite{Dandoy:2023jot, Franciolini:2023pbf, Franciolini:2023wjm}. Relieving the tension would require PBH production to be suppressed by specific non-gaussianities~\cite{Franciolini:2023pbf} or modifications to cosmology at the time of the QCD phase transition~\cite{Balaji:2023ehk}. 

\begin{figure}
    \centering
    \includegraphics[width=0.95\columnwidth]{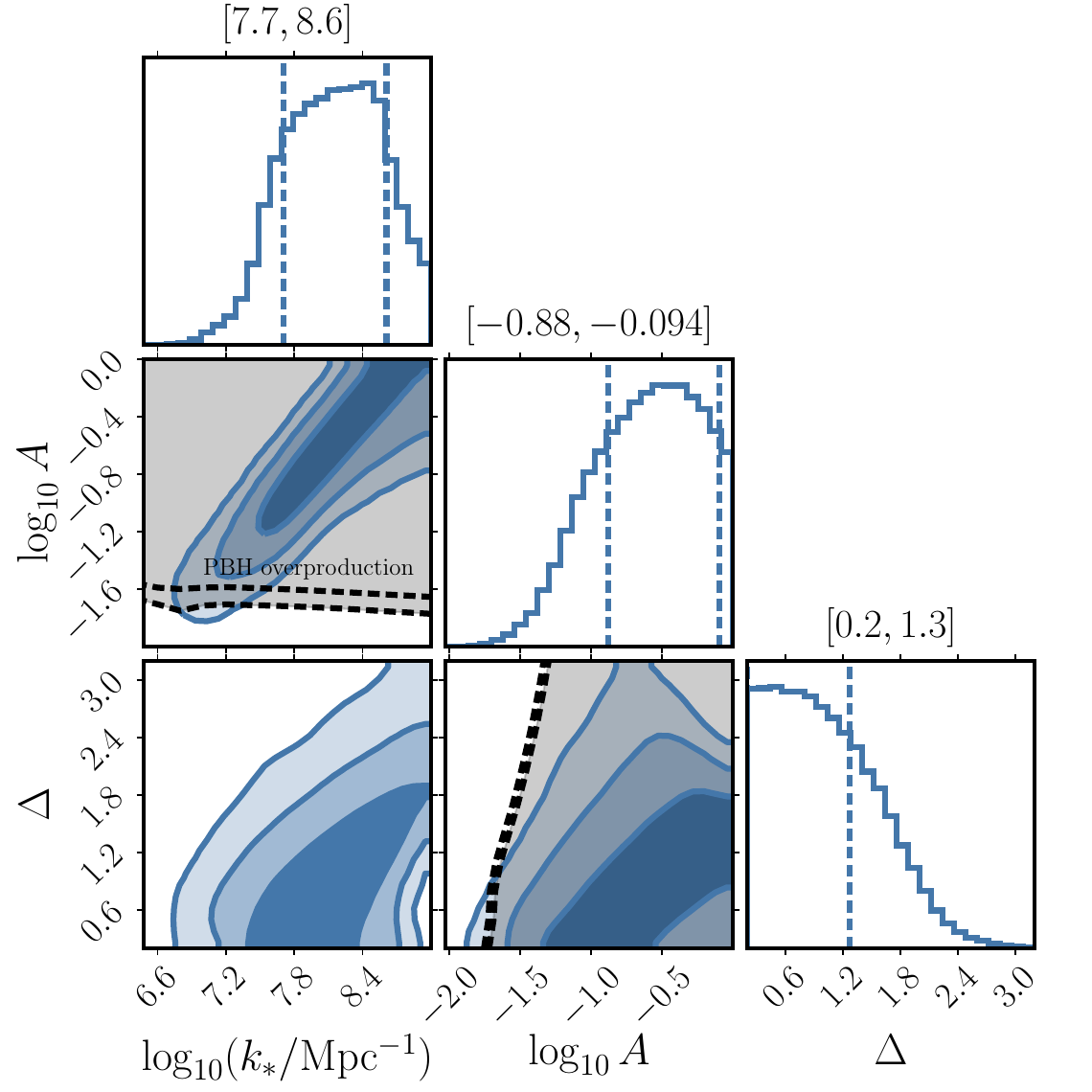}
    \caption{The posterior probability distribution of the SIGW fit to the NG15 data. The black shaded region indicates where $f_{\rm PBH}>1$, while the black dashed lines bracket the 1$\sigma$ ranges in the marginalised direction.}
    \label{fig:SIGWonly_posteriors}
\end{figure}

The present-day SIGW background would have a spectrum given by Eq.~\eqref{eq:OmegaGWtoday}, where~\cite{Inomata:2019yww,DeLuca:2019ufz,Yuan:2019fwv,Domenech:2020xin}
\begin{align}
\Omega_{\rm GW}(T) 
= \frac{1}{3} 
\int_1^\infty \td t \int_{-1}^1 \td s 
\frac{\overline{J^2(u,v)}}{(u v)^2} 
{\cal P}_\zeta (v k)
{\cal P}_\zeta (u k) ,
\label{eq:P_h_ts}
\end{align}
where $u=(t + s)/2$, $v=(t - s)/2$
and the transfer function $\overline{J^2 (u,v)}$ depends on the cosmological background~\cite{Kohri:2018awv,Espinosa:2018eve}.
The relevant calculations are reported in Appendix~\ref{app:NGSIGW}, including higher-order corrections to the SIGW signal that arise in the event of non-Gaussian curvature perturbations. 

A broad class of primordial curvature spectra can be characterized by a log-normal (LN) shape of the form
\be\label{eq:PLN}
    \mathcal{P}_{\zeta}(k)
    = \frac{A}{\sqrt{2\pi}\Delta} \, \exp\left( -\frac{1}{2\Delta^2} \ln^2(k/k_*) \right)\,.
\ee
The SIGW model therefore has three relevant parameters: the characteristic scale $k_*/{\rm Mpc}^{-1}$, the peak amplitude $A$, and the width $\Delta$.~\footnote{Similar results are obtained if one assumes a broken power-law power spectrum of the form inspired by inflection-point inflationary models. See Ref.~\cite{Franciolini:2023pbf} for an in-depth discussion.} The distributions of their posterior values and two-parameter correlations for fits to the NG15 data assuming only scalar-induced second-order GWs are shown in Fig.~\ref{fig:SIGWonly_posteriors}. The best-fit values of the SIGW model parameters are:
\be\label{eq:bestfitSIGW}
    \log_{10} (k_*/{\rm Mpc}^{-1})=7.7\,, \quad \log_{10} A=-1.2 \,, \quad \Delta=0.21 
\ee
for which $-2 \Delta \ell = -2.1$.

\begin{figure}
    \centering
    \includegraphics[width=1\columnwidth]{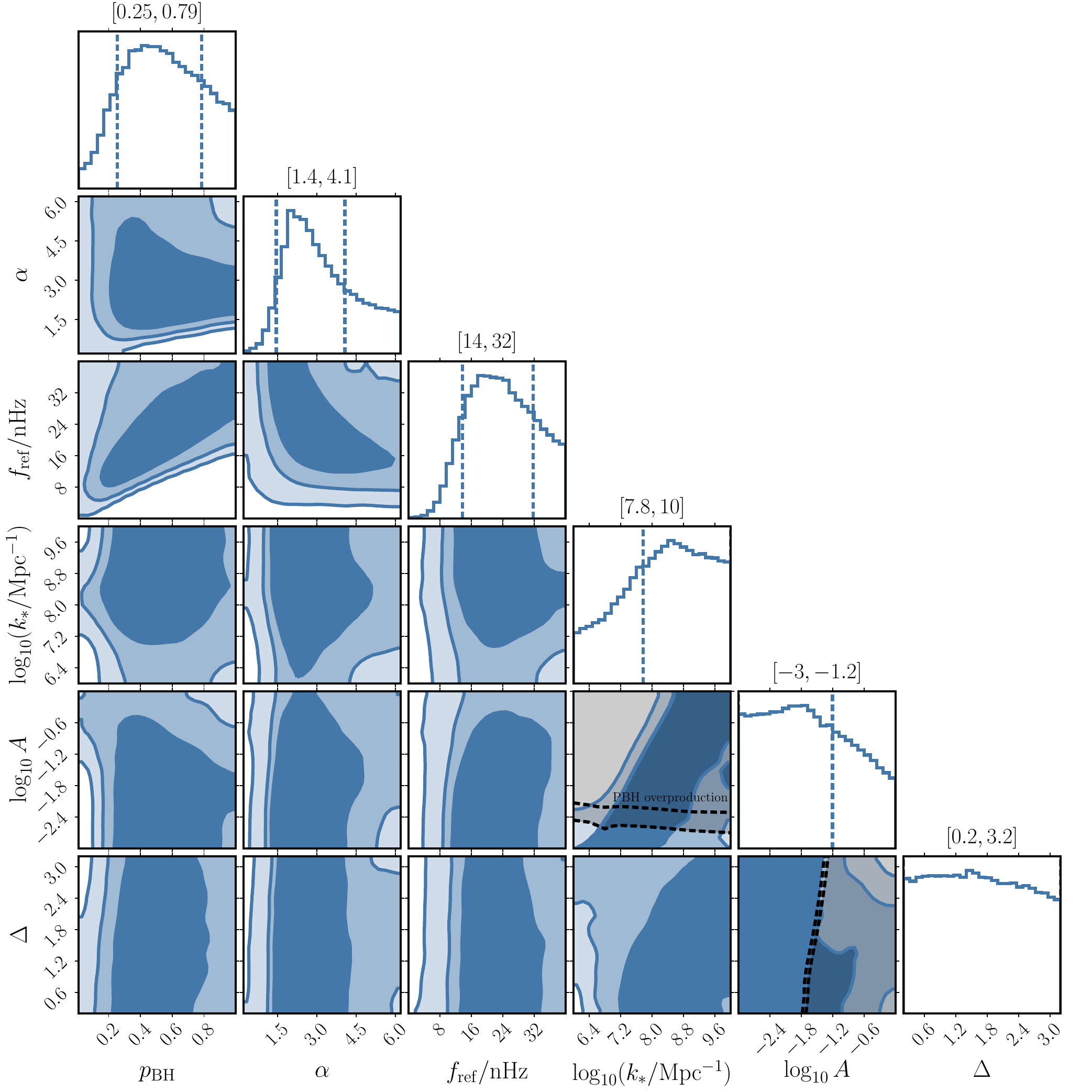}
    \caption{As in Fig.~\ref{fig:SIGWonly_posteriors} for a SIGW fit to NG15 data including also SMBH binaries with environmental effects.}
    \label{fig:SIGW_posteriors}
\end{figure}

We also show in Fig.~\ref{fig:SIGWonly_posteriors} the regions of parameter space where PBHs are overproduced. We compute their abundance following the prescription given in Refs.\,\cite{DeLuca:2019qsy,Young:2019yug,Germani:2019zez,Ferrante:2022mui}. In order to compute the abundance more precisely we also include the effects of the QCD phase transition and the shape of the power spectrum in the parameters, following Refs.\,\cite{Musco:2020jjb,Franciolini:2022tfm,Musco:2023dak}. As already pointed out in \cite{Franciolini:2023pbf}, the tension between the NANOGrav data and a scenario with the SIGW signal alone can be relaxed in models where large non-Gaussianities suppress the PBH abundance, or if the PBH formation take place during non standard cosmological phases deviating from radiation domination \cite{Balaji:2023ehk,Zhao:2023joc,Liu:2023pau}. Our results differ from those in Refs.\,\cite{Inomata:2023zup, Figueroa:2023zhu, Yi:2023mbm, Firouzjahi:2023lzg, You:2023rmn,Zhao:2023joc}. This discrepancy comes from several limitations of the analyses in these papers, such as the omission of critical collapse and the nonlinear relationship between curvature perturbations and density contrast, the adoption of a different value for
the threshold (independently from the curvature power spectrum), and the use of a different window function, without properly recomputing the threshold values \cite{Young:2019osy}.~\footnote{
When computing the PBH abundance in this work we have adopted the best prescription available following Ref.~\cite{Franciolini:2023pbf} and references therein. We mention that Ref.\,\cite{DeLuca:2023tun} (see, however, also \cite{Germani:2023ojx}) suggested that super-horizon threshold conditions may lead to an overestimation of the abundance, due to non-linear effects not included in the linear transfer function. }

Fig.~\ref{fig:SIGW_posteriors} shows a full set of posterior parameter distribution functions and two-parameter correlations for the fit to a combination of scalar-induced second-order GWs and SMBH binaries. The preferred values of the SIGW model parameters are similar to those in the SIGW-only fit, but their distributions are significantly broader, as was to be expected. The best-fit parameters of the combined model are
\bea
    &\log_{10} (k_*/{\rm Mpc}^{-1}) = 7.6 \,, \quad \log_{10} A = -1.2 \,, \quad \Delta = \, 0.30 \,, \\
    &p_{\rm BH}=0.06\,, \quad  f_{\rm ref}=26 \, {\rm nHz}\,, \quad \alpha=4.1\,.
\eea

The best-fit parameters describing the SIGW sector in the mixed scenario are very similar to those reported in Eq.~\eqref{eq:bestfitSIGW}, where SIGWs alone are assumed to explain PTA observations. This is because the SIGW model provides a better fit, and dominates the signal where the likelihood peaks. 
However, neither of the two models dominates over the other beyond the $\sim 2\sigma$ level, maintaining the large degeneracy of most parameters in the posterior distributions shown in Fig.~\ref{fig:SIGW_posteriors}.

Imposing a lower bound on $p_{\rm BH}$ forcing the SMBHs model to produce the candidate event at $4$nHz with probability larger than $5\%$, we find instead that the astrophysical channel is required to dominate, and the best fit parameters are
\bea
    &\log_{10} (k_*/{\rm Mpc}^{-1}) = 6.3 \,, 
    \quad \log_{10} A = -2.2 \,, 
    \quad \Delta = \, 0.21 \,, \\
    &p_{\rm BH}=0.70\,, 
    \quad  f_{\rm ref}=24 \, {\rm nHz}\,, 
    \quad \alpha=3.8\,.
\eea
with a maximum likelihood similar to the SMBHs-only scenario (with enviromental effects included), resulting in $-2 \Delta \ell=-0.4$. 

\subsection{First-order GWs}
\label{sec:FOGWs}

Another mechanism for generating a cosmological SGWB is during inflation via first-order GWs (FOGWs)~\cite{Grishchuk:1974ny,Starobinsky:1979ty}. We consider canonically normalized single-field slow-roll models, which predict a primordial tensor power spectrum that, in the relevant range of frequencies between the CMB and PTA scales, can be approximated by a pure power-law~\cite{NANOGrav:2023hvm}:
\be \label{eq:FO}
    \Omega_{\rm GW}(f) = \frac{r A_s }{24}\left(\frac{f}{f_{\rm CMB}}\right)^{n_t}\mathcal{T}(f) \, ,
\ee
where $f_{\rm CMB} = 7.7\times 10^{-17}$\,Hz, $A_s = 2.1\times 10^{-9}$~\cite{Planck:2018vyg}, $n_t$ is the tensor spectral index and $r$ is the tensor-to-scalar ratio. The transfer function that connects the radiation-dominated era to reheating can be approximated as~\cite{Kuroyanagi:2014nba,Kuroyanagi:2020sfw}
\be
    \mathcal{T}(f) \approx \frac{\Theta\left(f_{\rm {end }}-f\right)}{1-0.22\left(f / f_{\mathrm{rh}}\right)^{1.5}+0.65\left(f / f_{\mathrm{rh}}\right)^2} \,,
\ee
where $f_{\rm end}$ and $f_{\rm rh} \leq f_{\rm end}$ correspond, respectively, to the end of inflation and to the end of reheating. For simplicity, we assume that the cut-off scale is $f_{\rm end} \gg f_{\rm rh} = f_{H}(T_{\rm rh})$. In this way, the analysis is independent of the choice of the cutoff. 

\begin{figure}
    \centering
    \includegraphics[width=0.95\columnwidth]{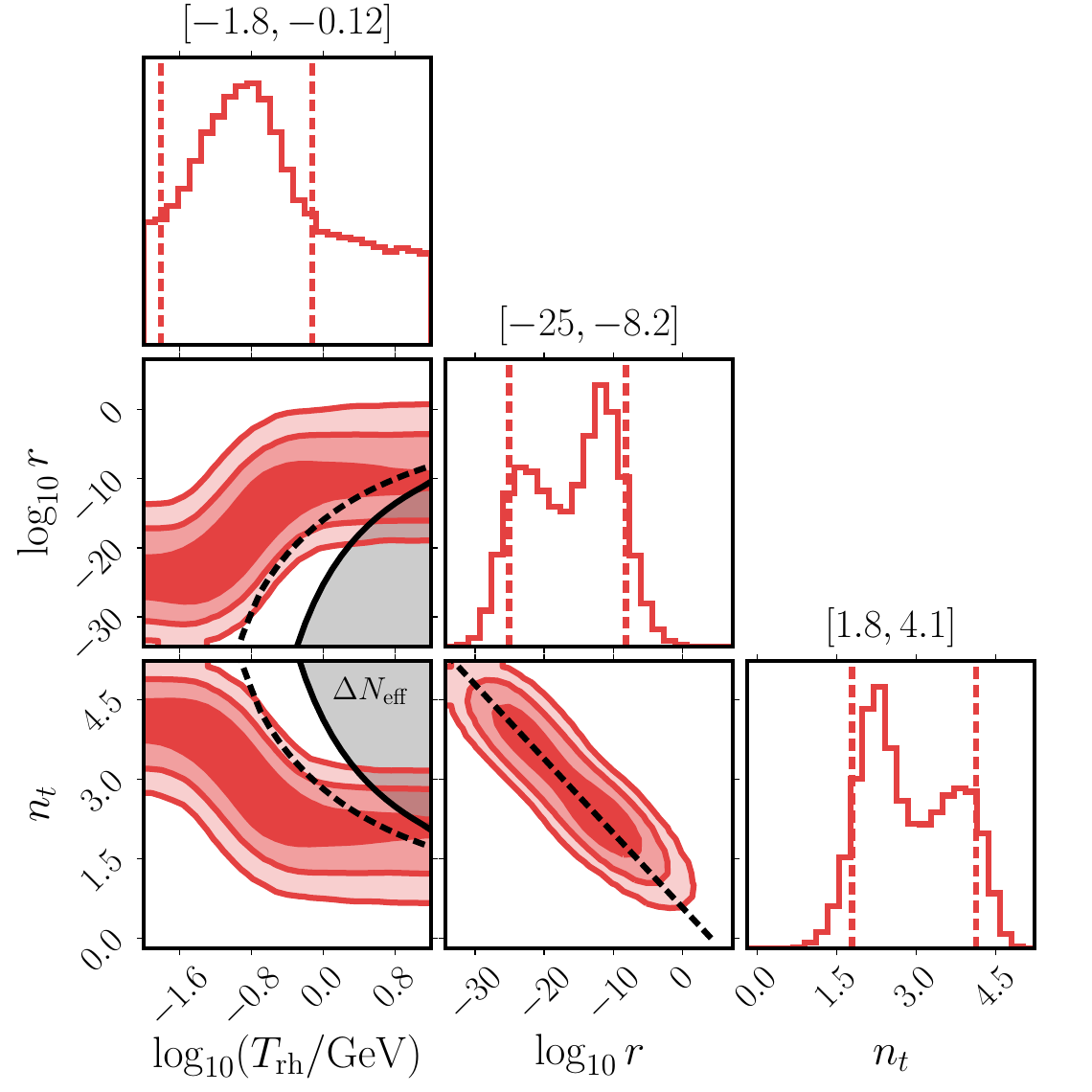}
    \caption{ The posterior probability distribution of the FOGW fit to the NG15 data. The black-shaded region shows the constraints by $\Delta N_{\rm eff}$  bound in Eq.~\eqref{eq:deltaN}, assuming $f_{\rm end} = f_{\rm rh}$ (solid line) and $f_{\rm end} =10 f_{\rm rh}$ (dashed line). We also show the $r-n_t$ correlation \eqref{eq:nt-rcorr} in the corresponding panel. }
    \label{fig:FOGWonly_posteriors}
\end{figure}

The FOGW scenario is parametrized by the inflationary parameters $n_t$ and $r$, and the reheating temperature $T_{\rm rh}$. As seen in the left panel of Fig.~\ref{fig:FOGWonly_posteriors}, their values are very tightly correlated, and their best-fit values are  
\be\label{eq:figwalone_bestfit}
    T_{\rm rh} = -0.67\,{\rm GeV} \,, 
    \quad  \log_{10} r = -14 \,, 
    \quad  n_{t} \, = \, 2.6
\ee
for which  $-2 \Delta \ell=-2.0$. 

\begin{figure}
    \centering
    \includegraphics[width=1\columnwidth]{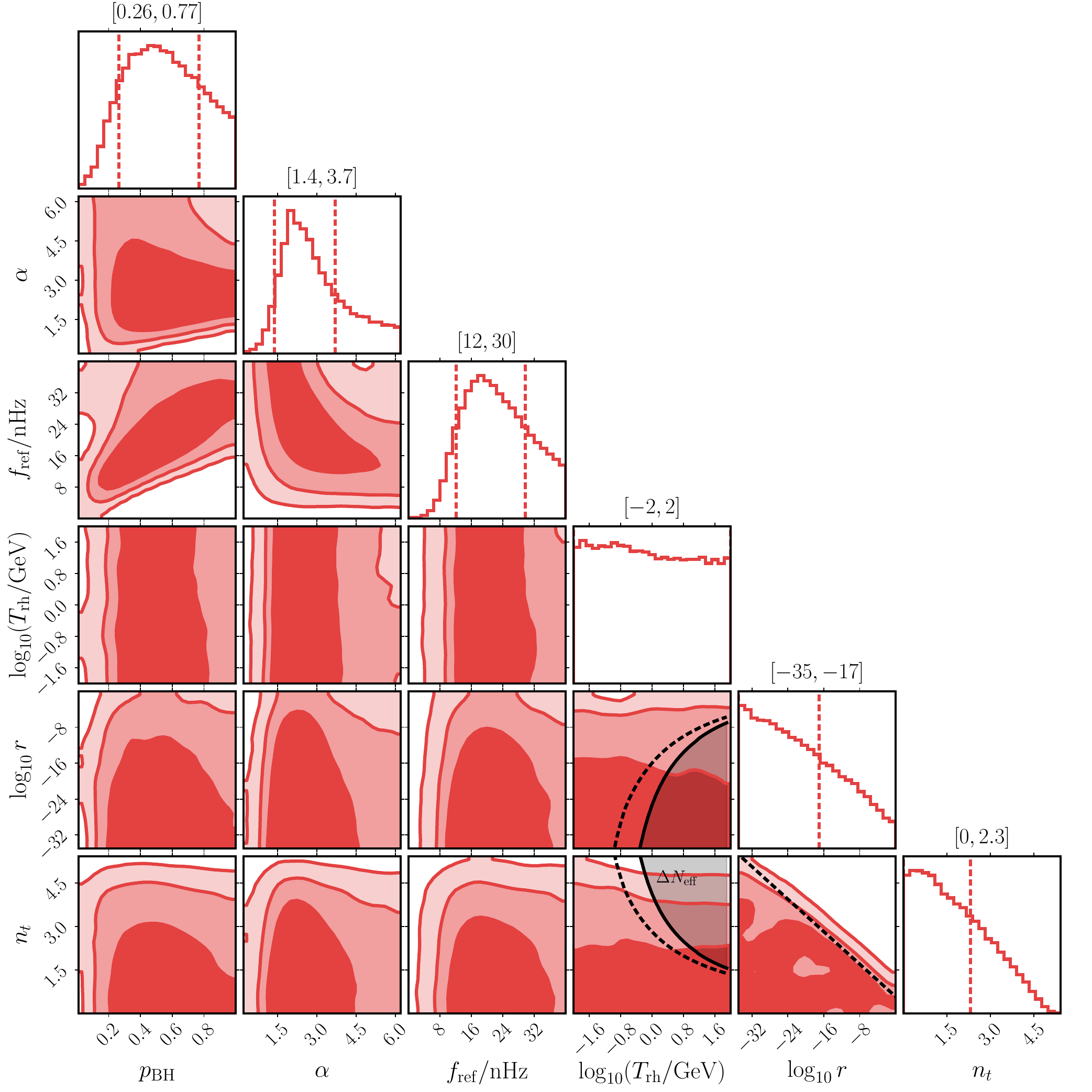}
    \caption{Same as Fig.~\ref{fig:FOGWonly_posteriors} for a FOGW fit to NG15 data including also SMBH binaries with environmental effects.}
    \label{fig:FOGWSMBH_posteriors}
\end{figure}

A constraint on the possible values of these parameters is given by the SGWB contribution to the radiation energy budget characterized by $\Delta N_{\rm eff}$, i.e. its contribution to the effective number of relativistic species. $\Delta N_{\rm eff}$ is constrained by BBN and CMB probes, and an approximate upper limit is given by $\Delta N_{\rm eff} \lesssim0.3$~\cite{Planck:2018vyg,Fields:2019pfx}. The GW spectrum, integrated from BBN scales to the cutoff frequency $f_{\rm end}$, must not exceed an upper limit that is set by the allowed amount of extra relativistic degrees of freedom at the time of BBN and recombination. This contribution is given~\cite{Smith:2006nka,Boyle:2007zx}
\be \label{eq:deltaN}
    1.8\times 10^{5}\int_{f_{\mathrm{BBN}}}^{f_{\mathrm{end}}} \frac{\mathrm{d} f}{f}  \Omega_{\mathrm{GW}}(f)h^2 \lesssim  \Delta N_{\mathrm{eff}}^{\max }
\ee
where we set $f_{\rm BBN}\simeq10^{-12}$~Hz, which is the frequency at horizon re-entry of tensor mode around the onset of BBN
at $T \simeq 10^{-4}$ GeV\,\cite{Caprini:2018mtu}. 
For a slope compatible with the NANOGrav data,  the signal cannot extend to excessively high frequencies, as doing so would violate constraints imposed by BBN\,\cite{Maggiore:1999vm,Pagano:2015hma}. For these reasons, we limit our analysis to the range $T_{\rm rh}\in [0.01,10]$ ${\rm GeV}$. As can be seen from Fig.\ref{fig:FOGWonly_posteriors}, if the reheating temperature is $\lesssim 0.3$ GeV the preferred signal changes the tilt moving from $n_t$ to $n_t-2$. This is because, for such values of the temperature, the cut-off of the signal is at a lower frequency than the last NANOGrav bin, and the GW spectrum in the PTA band is composed of tensor modes that re-entered the horizon during reheating after inflation.
When deriving the constraint from the bound on $\Delta N_{\rm eff}$, we find that $n_t$ and $r$ are strongly correlated, and the equation that relates these quantities can be approximated by
\be \label{eq:nt-rcorr}
    n_t = -0.14\log_{10} r +0.58.
\ee
Using this last relation we constrain the parameter space as shown in Figs.~\ref{fig:FOGWonly_posteriors} and \ref{fig:FOGWSMBH_posteriors}.

Our results are in perfect agreement with Refs.~\cite{NANOGrav:2023hvm, Vagnozzi:2023lwo}.
Fig.~\ref{fig:FOGWSMBH_posteriors} shows a full set of posterior parameter distribution functions and three-parameter correlations for the fit to a combination of FOGWs and SMBH binaries. 
The best fit of the combined models results in
 $-2 \Delta \ell=-1.9$ with $p_\text{BH}\simeq 0$.
 Thus, the FIGW contribution is dominant, and the best-fit parameter of the FIGW sector coincides with those reported in Eq.~\eqref{eq:figwalone_bestfit}, assuming no SMBH contribution. 
Imposing a lower bound on $p_{\rm BH}$ to produce the candidate SMBH merger at $4$nHz, 
we find instead
\bea
    &T_{\rm rh} = 1.5\,{\rm GeV} \,, \quad \log_{10} r = -9.1 \,, \quad n_{t} \, = \, 1.46 \, , \\
    &p_{\rm BH}=0.35\, , \quad  f_{\rm ref}=16 \, {\rm nHz} \,, \quad \alpha=4.7\, .
\eea
with resulting in $-2 \Delta \ell=-0.1$.

\subsection{``Audible" axions}
\label{sec:axions}

\begin{figure}
    \centering
    \includegraphics[width=0.91\columnwidth]{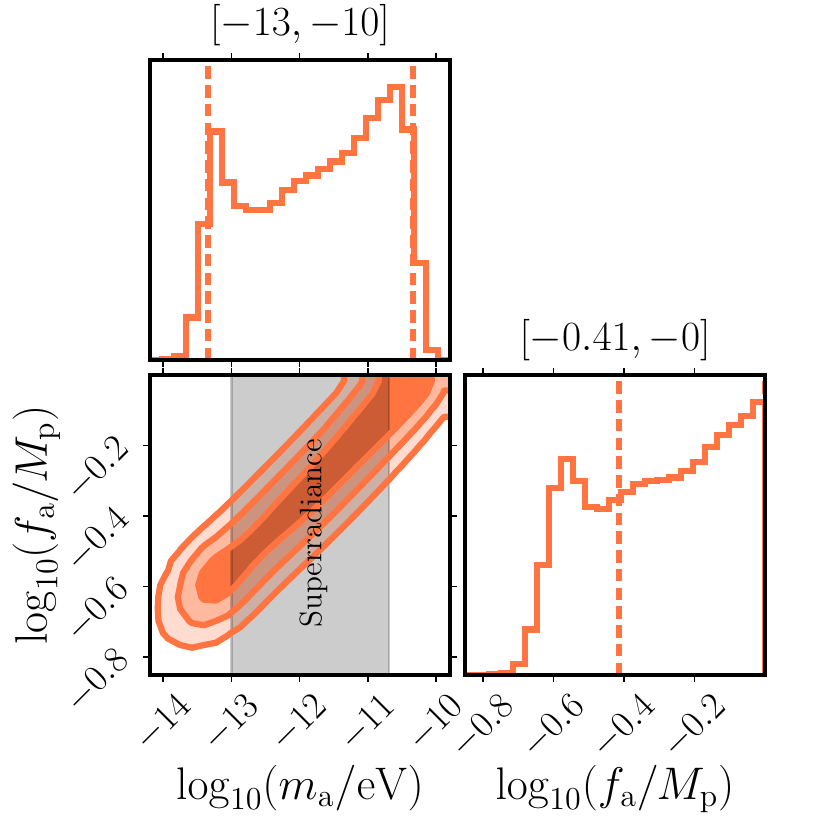} 
    \caption{The posterior probability distribution of the ``audible" axion fit to the NG15 data. The shaded region is excluded by super-radiance constraints.}
    \label{fig:Axion_posteriors}
\end{figure}

In models in which an axion or an axion-like particle has a weak coupling to the SM, thanks to a coupling to a light dark photon, a SGWB can be produced~\cite{Machado:2018nqk, Machado:2019xuc,Co:2021rhi,Fonseca:2019ypl,Chatrchyan:2020pzh,Ratzinger:2020oct,Eroncel:2022vjg}.
After the axion begins to roll, it induces a tachyonic instability for one of the dark photon helicities, causing vacuum fluctuations to grow exponentially. This effect generates time-dependent anisotropic stress in the energy-momentum tensor, which ultimately sources GWs. In this scenario, the condition of having a coupling with a dark photon and not with the SM photon is due to the fact that the latter undergoes rapid thermalization that would destroy the conditions required for exponential particle production.\footnote{GWs can also be produced in kinetic misalignment scenarios (in which the axion field has a nonzero initial velocity) also in the absence of dark photons. As shown in \cite{Machado:2019xuc} the resulting signal in this scenario is, however, in general too small to be observable in the foreseeable future.} The GW formation ends when the tachyonic band closes at temperature~\cite{Machado:2018nqk}
\be
    T_* \approx \frac{1.2 \sqrt{m_a M_{\rm P}}}{g_*^{1/4} (\alpha \theta)^{2/3}} \,,
\ee
where $\alpha$ is the coupling with the dark photon, $\theta$ is the initial misalignment angle, and $m_a$ is the mass of the axion.
As shown in~\cite{Machado:2019xuc}, assuming that between the beginning of the axion oscillations and the end of the GW formation, the effective number of degrees of freedom does not change, the GW spectrum at temperature $T_*$ can be fitted by
\be \label{eq:OmegaAa}
    \Omega_{\mathrm{GW}}(f,T_*) = \frac{6.3 \left(\frac{f_a}{M_{\rm P}}\right)^4\left(\frac{\theta^2}{\alpha}\right)^{4/3} S_H(f,f_H(T_*))}{\left[\frac{f}{2.0 f_p}\right]^{\!-1.5}+\exp\!\left[12.9\left(\frac{f}{2.0 f_p}-1\right)\right]} \,,
\ee
where $f_a$ is the decay constant of the axion and the peak frequency of the spectrum is
\be
    f_p \approx 2.5 (\alpha \theta)^{4/3} f_H(T_*) \,.
\ee 
Compared to the fit in~\cite{Machado:2019xuc}, we have added the function $S_H(f,f_H)$ that accounts for the causality tail of the spectrum at frequencies $f<f_H(T_*)$ and is given by Eq.~\eqref{eq:SH} with $a=1.5$. We set the parameter $\delta$ in $S_H(f,f_H)$ to $\delta = 1$. Typical values for the initial misalignment angle are of $\theta\sim\mathcal{O}(1)$ \,\cite{DiLuzio:2020wdo,Marsh:2015xka} and Eq.~\eqref{eq:OmegaAa} holds only for $\theta \sim 1$ and $\alpha \geq 10$, the range of values where the particle production process is efficient \cite{Agrawal:2017eqm}. For simplicity, we fix $\theta=1$  and $\alpha=20$ in the analysis. 

\begin{figure}
    \centering
    \includegraphics[width=1\columnwidth]{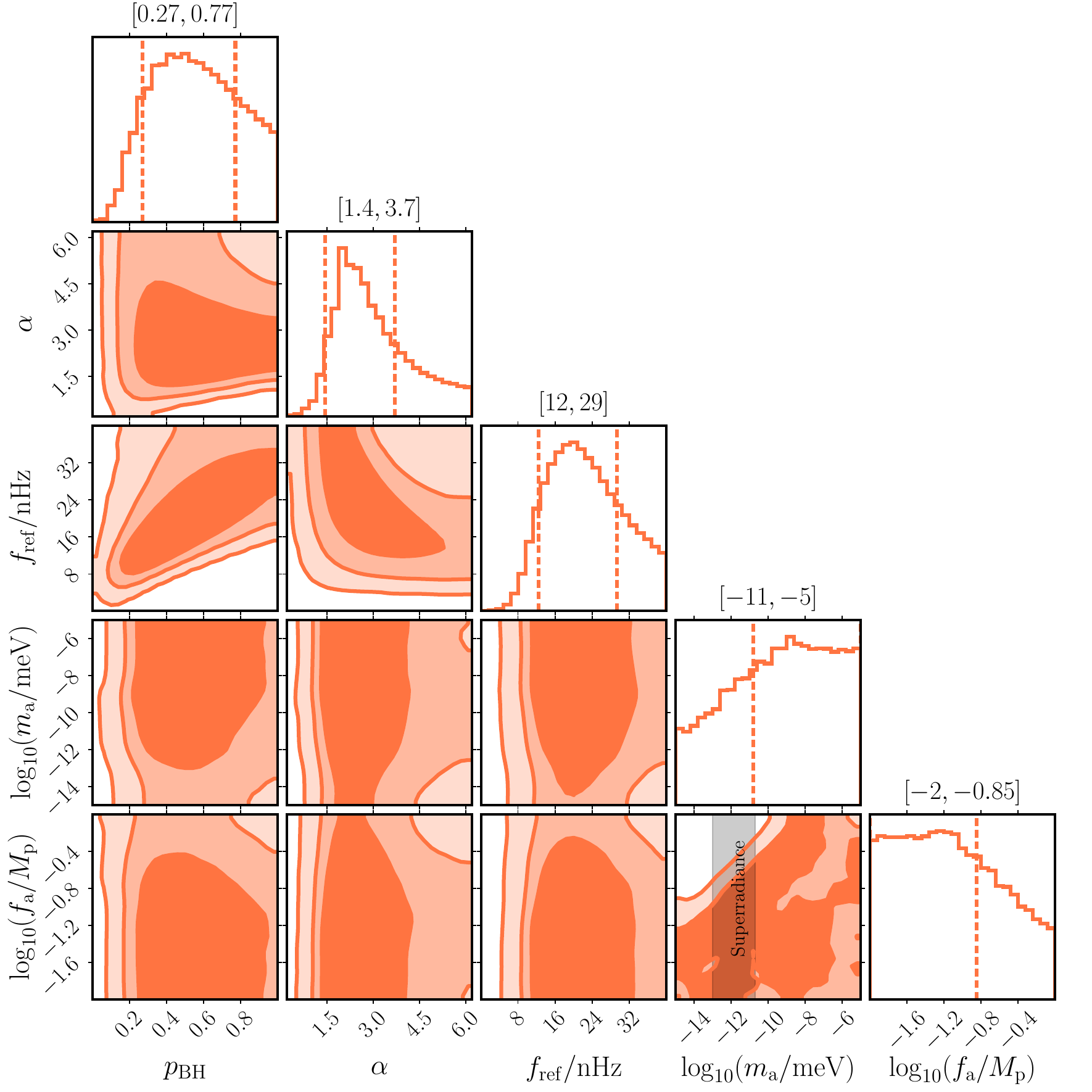}
    \caption{As in Fig.~\ref{fig:Axion_posteriors}  for an ``audible" axion fit to NG15 data including also SMBH binaries with environmental effects.}
    \label{fig:aaSMBH_posteriors}
\end{figure}

\begin{figure*}
\centering
\includegraphics[width=0.85\textwidth]{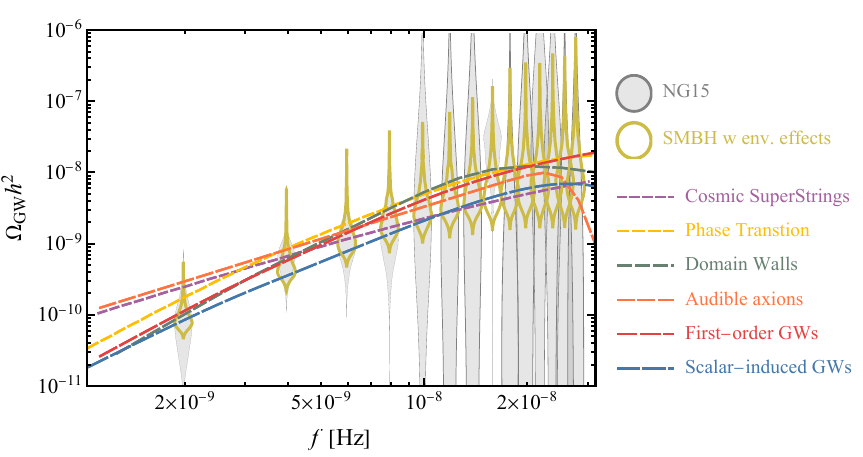}
\caption{Comparison of the best fits to the NG15 data for SMBH binaries with environmental effects and in the indicated BSM cosmological models.}
\label{fig:Bestfits}
\end{figure*}

We show in Fig.~\ref{fig:Axion_posteriors} the posteriors of the ``audible" axions two-parameter fit to the NG15 data. The best-fit values are achieved at
\be
    m_a= 3.1\times10^{-11} {\rm eV}\,,\quad f_a  =0.87\, M_{\rm P} \,,
\ee
corresponding to $-2 \Delta \ell = -1.6$ relative to the baseline. Differently from the analysis performed in ref.\,\cite{Figueroa:2023zhu} we find a larger possible range for the masses of the axions below super-Planckian $f_a$ values. Part of this region is currently ruled out by the super-radiance constraints\,\cite{Zhang:2021mks,Baryakhtar:2020gao,Mehta:2020kwu}. { The discrepancy between our analysis and that in \cite{Figueroa:2023zhu} is twofold. First, the analysis in Ref.\,\cite{Figueroa:2023zhu} is based on both NANOGrav and EPTA data, with the latter dataset favoring larger amplitudes at higher frequencies. Heavier axions move the main peak to higher frequencies without changing the amplitude of the main peak. As a consequence, including the EPTA dataset leads to stronger bounds on more massive axions compared to our analysis. Lastly, when the peak of the signal is at frequencies much higher than the PTA range, which occurs for the larger masses in Fig.\ref{fig:Axion_posteriors}, the causality tail dominates the signal compared to the Ansatz used to characterize the signal in \cite{Figueroa:2023zhu}}.

Fig.~\ref{fig:aaSMBH_posteriors} shows a full set of posterior parameter distribution functions and two-parameter correlations for the fit to a combination of ``audible" axions and SMBH binaries. The best fit of the combined fit is at the same point as without the SMBH, $p_{\rm BH} \approx 0$, so the inclusion of the SMBH binaries does not improve the fit. If we impose the posteriors from the candidate event at $4{\rm nHz}$, then the best-fit shifts to 
\bea 
    &p_{\rm BH}=0.58\, , \; f_{\rm ref}=20 \, {\rm nHz}\, , \; \alpha=5\, ,
\eea
and same axion parameters, corresponding to $-2 \Delta \ell = -0.3$.

\section{Model Comparisons} 
\label{sec:comp}

Fig.~\ref{fig:Bestfits} compares the best fits in the models we have studied in the frequency range measured by NANOGrav. We see that the cosmological models all capture to some extent the frequency dependence measured in the NANOGrav range, as does the astrophysical SMBH binary scenario if environmental effects are included. This model, the cosmic (super)string scenario and ``audible" axions fit the data less well than phase transitions, domain walls, scalar-induced GWs and first-order GWs, as seen quantitatively in Table~\ref{tab:results}, which collects the numerical results for our best fits in the astrophysical SMBH model with and without environmental effects and in the BSM cosmological scenarios we have studied. The best-fit values of the model parameters are shown in the second column, and the values of the Bayesian inference criterion $\Delta {\rm BIC}$ relative to the baseline SMBH model with the environmental effects are shown in the third column. The numbers in the parenthesis in the third column indicate the Bayesian inference criterion of the fit combining the cosmological model with the SMBH model including the environmental effects. These fits give significantly larger values of $\Delta {\rm BIC}$ because they include more parameters. The combined fit is better than the SMBH model alone only in the case of phase transitions and domain walls.

\begin{figure*}
\centering
\includegraphics[width=0.95\textwidth]{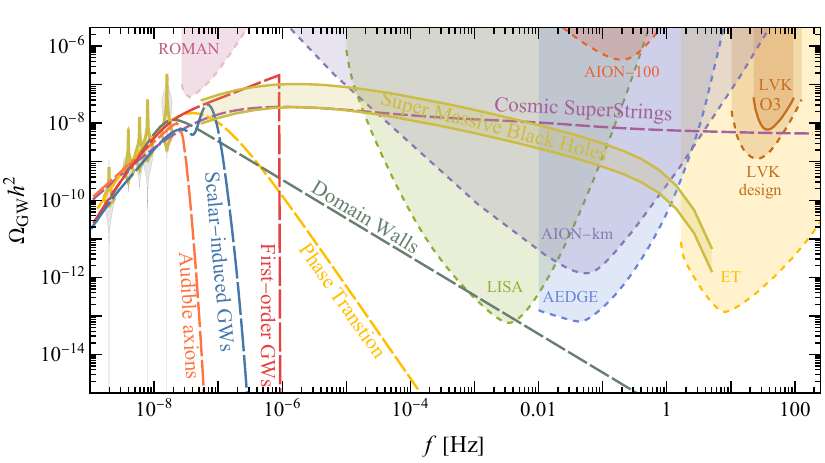}
\caption{Extension of Fig.~\ref{fig:Bestfits} to higher frequencies, indicating the prospective sensitivities of LVK and planned and proposed future detectors. For clarity, we include only the first four and the eighth bins which are the most narrow and have the largest impact on fit quality. 
The green band extends the ``violins" in the PTA  range to higher frequencies, and shows the mean GW energy density spectrum from SMBH binaries heavier than $10^3 M_\odot$ for $p_{\rm BH} = 0.25 - 1$. Individual SMBH binaries are expected to be measurable in this frequency range. The FOGW spectrum is expected to be cut off at some frequency between the ranges where PTAs and interferometers are sensitive.}
\label{fig:BestFits2}
\end{figure*}

Finally, the fifth fourth of Table~\ref{tab:results} tabulates some prospective signatures of the models considered that could help discriminate between them. As discussed above, PTAs should be sensitive to bin-to-bin fluctuations if the observed SGWB is due to SMBH binaries. Moreover, there should be observable anisotropies in the SGWB~\cite{Sato-Polito:2023spo} and circular polarization could be observable~\cite{Sato-Polito:2021efu, Ellis:2023owy}. Also, future PTA data could reveal individual binary sources. We see from the fits combining the cosmological models with the SMBH model with environmental effects that the 1$\sigma$ ranges for the SMBH parameters are relatively broad. A consequence of this is that it is possible to find a mixed scenario where the fit is better than in the pure SMBH model but that can still accommodate observable individual binary sources.

Fig.~\ref{fig:BestFits2} extends the model comparisons to the higher frequencies accessible to other experiments including LISA~\cite{Bartolo:2016ami, Caprini:2019pxz, LISACosmologyWorkingGroup:2022jok}, The Einstein Telescope (ET)~\cite{Punturo:2010zz, Hild:2010id}, AION~\cite{Badurina:2019hst,Badurina:2021rgt}, AEDGE~\cite{AEDGE:2019nxb,Badurina:2021rgt}, the Nancy Roman telescope (ROMAN)~\cite{Wang:2022sxn} and the design sensitivity of LVK~\cite{LIGOScientific:2014pky, LIGOScientific:2016fpe, LIGOScientific:2019vic}. 
We see that LISA should be able to detect individual SMBH binaries, albeit with lower masses (${\cal O}(10^6 - 10^9)$ solar masses) than those responsible for the NANOGrav signal ($\gtrsim {\cal O}(10^9)$ solar masses~\cite{Ellis:2023owy}), at larger rates if environmental effects are important~\cite{Ellis:2023dgf}. Similarly, proposed experiments in the mid-frequency band such as AION and AEDGE would be sensitive to mergers of BHs with masses ${\cal O}(10^3 - 10^6)$ solar masses. On the other hand experiments such as LVK and ET are not sensitive to mergers of BHs with masses $\gtrsim 10^3$ solar masses.  In contrast, all these experiments should be able to detect GWs from cosmic (super)strings, even in the case of modified cosmological evolution~\cite{Ellis:2023tsl}.  On the other hand, the SGWB from a cosmological phase transition, SIGWs, FOGWs or ``audible" axions would be unobservable in any higher-frequency detector, whereas the SGWB from domain walls might be observable by LISA between $10^{-3} - 10^{-2}$~Hz, but probably not by AEDGE and other detectors operating at frequencies $\gtrsim 10^{-2}$~Hz.

\renewcommand{\arraystretch}{1.2}
\setlength{\tabcolsep}{3pt}
\begin{table*}[t]
\begin{center}
\textbf{Results from Multi-Model Analysis (MMA)}\\
\vspace{1mm}
\begin{tabular}{|p{0.28\textwidth}|p{0.2\textwidth}|c|l|}
\hline	
Scenario & Best-fit parameters & $\Delta {\rm BIC}$ & Signatures \\
\hline \hline
GW-driven SMBH binaries & $p_{\rm BH} = 0.07$ & 6.0 & FAPS, LISA, mid-$f$, \sout{LVK, ET} \\
\hline
GW + environment-driven & $p_{\rm BH} = 0.84$ & Baseline & FAPS, LISA, mid-$f$, \sout{LVK, ET} \\
SMBH binaries  & $\alpha = 2.0$ & (${\rm BIC} = 53.9$) & \\
 & $f_{\rm ref} = 34$~nHz & & \\
 \hline \hline
Cosmic (super)strings & $G \mu =2\times 10^{-12}$ & -1.2 &\sout{FAPS}, LISA, mid-$f$, {LVK, ET} \\
(CS) & $p = 6.3\times10^{-3}$ & (4.6) & \\
 \hline
Phase transition & $T_* = 0.34$~GeV & -4.9 &\sout{FAPS, LISA, mid-$f$, LVK, ET}\\
(PT) & $\beta/H = 6.0$ & (2.9) & \\
\hline
Domain walls & $T_{\rm ann} = 0.85$~GeV & -5.7 & \sout{FAPS}, LISA?, \sout{mid-$f$, LVK, ET}\\
(DWs) & $\alpha_* = 0.11$ & (2.2) & \\
 \hline
Scalar-induced GWs & $k_* = 10^{7.7}/{\rm Mpc}$ & -2.1 &
\sout{FAPS, LISA, mid-$f$, LVK, ET}\\
(SIGWs) & $A = 0.06$ & (5.8) & \\
& $\Delta = 0.21$ & & \\
\hline
First-order GWs & $\log_{10}{r}=-14$ & -2.0 & \sout{FAPS, LISA, mid-$f$, LVK, ET} \\
(FOGWs) & $n_{\rm t}= 2.6$ & (6.0) & \\
 & $T_{\rm rh} = -0.67$\,GeV &  & \\
\hline
``Audible" axions & $m_{a}= 3.1 \times 10^{-11}\, {\rm eV}$ & -4.2 & \sout{FAPS, LISA, mid-$f$, LVK, ET} \\
 & ${f_a}= 0.87\, M_{\rm P}$ & (3.7) & \\
\hline
\end{tabular} \\
\vspace{1mm}
{\footnotesize FAPS~$\equiv$~fluctuations, anisotropies, polarization, sources, mid-$f$~$\equiv$~mid-frequency experiment, e.g., AION~\cite{Badurina:2019hst}, AEDGE~\cite{AEDGE:2019nxb},\\ LVK~$\equiv$~LIGO/Virgo/KAGRA~\cite{LIGOScientific:2014pky,LIGOScientific:2016fpe,LIGOScientific:2019vic}, ET~$\equiv$~Einstein Telescope~\cite{Sathyaprakash:2012jk} (or Cosmic Explorer~\cite{Reitze:2019iox}), \sout{signature}~$\equiv$~not detectable}
\end{center}
\vspace{-4mm}
\caption{\it 
The parameters of the different models are defined in the text. For each model, we tabulate their best-fit values, and the Bayesian information criterion $BIC\equiv -2 \ell + k \ln 14$, where $k$ denotes the number of parameters, relative to that for the purely SMBH model with environmental effects that we take as the baseline. The quantity in the parentheses in the third column shows the $\Delta$BIC for the best-fit combined SMBH+cosmological scenario. The last column summarizes the prospective signatures.}
\label{tab:results}
\end{table*}

\section{Conclusions} 
\label{sec:concl}

The recent PTA data from NANOGrav and other PTA Collaborations have established the existence of a SGWB in the nHz range, but its interpretation remains an open question. Binary SMBH systems are the default astrophysical interpretation but remain to be established as the sources of the PTA signals. Many cosmological models invoking generic aspects of BSM physics have also been proposed as prospective sources. We have presented in this paper a comprehensive Multi-Model Analysis (MMA) that applies a common approach to assess the relative qualities of fits in these models, both with and without the inclusion of a SMBH binary background. We find that these models are capable of fitting the NANOGrav data at least as well as SMBH binaries alone (significantly better if environmental effects on the evolution of the binaries can be neglected). Future PTA datasets may be able to distinguish between the cosmological and astrophysical scenarios, e.g., by observing bin-to-bin fluctuations, anisotropies, circular polarization or individual sources. Observations at higher frequencies will also be invaluable: some cosmological scenarios predict a turndown of the frequency spectrum that precludes measurements by LISA, whereas astrophysical models suggest that BH mergers may be observable by LISA and mid-frequency experiments, and cosmic (super)strings predict that a SGWB background should be detectable by LISA and higher-frequency experiments.\\
~~\\
\indent
The discovery of nHz GWs may be linked to the biggest bangs since the Big Bang, namely mergers of SMBHs, or be providing evidence for new fundamental physics that is inaccessible to laboratory experiments. Time and more data will tell.

\begin{acknowledgments}
The work of J.E. was supported by the United Kingdom STFC Grants ST/T000759/1 and ST/T00679X/1.
G.F. acknowledges the financial support provided under the European Union's H2020 ERC, Starting Grant agreement no.~DarkGRA--757480 and under the MIUR PRIN programme, and support from the Amaldi Research Center funded by the MIUR program ``Dipartimento di Eccellenza" (CUP:~B81I18001170001).
This work was supported by the EU Horizon 2020 Research and Innovation Programme under the Marie Sklodowska-Curie Grant Agreement No. 101007855 and additional financial support provided by ``Progetti per Avvio alla Ricerca - Tipo 2", protocol number AR2221816C515921.
M.L. acknowledges the financial support provided by the Polish Returns Programme under agreement PPN/PPO/2020/1/00013/U/00001 and
the Polish National Science Center grant 2018/31/D/ST2/02048.
The work of G.H., V.V. and H.V. was supported by the European Regional Development Fund through the CoE program grant TK133 and by the Estonian Research Council grants PRG803 and PSG869. The work of V.V. has been partially supported by the European Union's Horizon Europe research and innovation program under the Marie Sk\l{}odowska-Curie grant agreement No. 101065736.
A.J.I. acknowledges the financial support provided under the ``Progetti per Avvio alla Ricerca Tipo 1", protocol number AR12218167D66D36, and the ``Progetti di mobilità di studenti di dottorato di ricerca". 
\end{acknowledgments}

\newpage
\onecolumngrid
\appendix

\section{Second-order induced gravitational waves with non-Gaussianities}\label{app:NGSIGW}

We summarize here the formulas used to compute the spectrum of SIGWs when non-Gaussianities are included, referring the reader to Refs.~\cite{Unal:2018yaa, Cai:2018dig, Yuan:2020iwf, Adshead:2021hnm, Abe:2022xur, Chang:2022nzu, Garcia-Saenz:2022tzu, Li:2023qua} for more details. 
We assume that the curvature perturbation can be written in terms of its Gaussian component, as 
$
\zeta = \zeta_{\rm G} + F_{\rm NL}\zeta_{\rm G}^2 .
$
The non-linear parameter is often normalised as $F_{\rm NL } = (3/5) f_{\rm NL} $.

The higher-order corrections to the GW power spectrum can then be written as
\begin{equation}\label{eq:Omegabar-total}
   \Omega_{\rm GW}(T) 
    =\Omega_{\rm GW}^{(0)} (T)
    +\Omega_{\rm GW}^{(1)} (T)
    +\Omega_{\rm GW} ^{(2)} (T) \ ,
\end{equation}
where each contribution scales as
${\cal P}_h (k,\eta) ^{(n)} (\eta,k) \propto A^2 (A F_\text{NL}^2)^n$ and the first order is given in Eq.~\eqref{eq:P_h_ts}. 
The energy-density fraction spectrum of SIGW at the current epoch is then given by Eq.~\eqref{eq:OmegaGWtoday}.

\begin{figure}[ht!]
    \begin{center}
    \includegraphics[width=0.52\columnwidth]{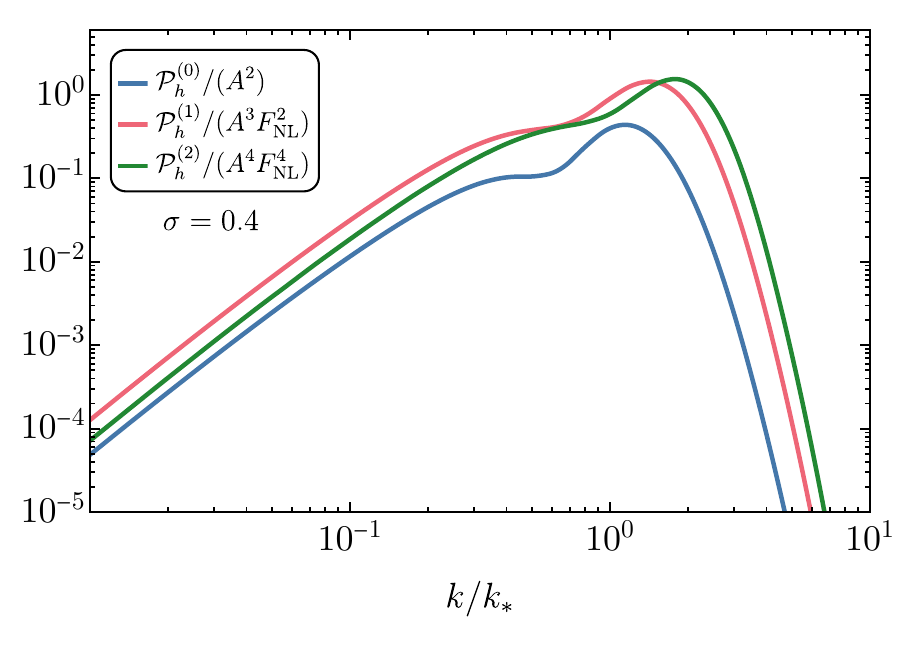}
    \end{center}
    \vspace{-1cm}
    \caption{Contributions to the SIGW spectrum from the leading orders in perturbation theory, where we have factored out the relevant combination of the amplitude $A$ and non-Gaussianity $F_\text{NL}$.
    This plot assumes a log-normal curvature power spectrum centered at $k_*$ that is characterised by a width $\sigma = 0.4$.}
    \label{fig:SIGW-fNL_spectra}
\end{figure}

In accordance with Refs.~\cite{Adshead:2021hnm, Li:2023qua, Zhu:2023faa, Ragavendra:2021qdu, Atal:2021jyo}, the higher-order terms in \eqref{eq:Omegabar-total} can be expressed as
\begin{align}
\Omega_{\rm GW}^{(1)} (T)
=& \frac{2F_{\rm NL}^2}{3} 
\prod_{i=1}^2 
\biggl[\int_1^\infty \td t_i \int_{-1}^1 \td s_i\, \frac{16}{(t_i^2-s_i^2)^2}\biggr] 
\Bigg\{ \frac{1}{2}
\overline{J^2 (s_1,t_1)} 
{\cal P}_\zeta  (v_1 v_2 k) 
{\cal P}_\zeta  (u_1 k) 
{\cal P}_\zeta  (v_1 u_2 k)
\nonumber
\\
+& \int_0^{2\pi} \frac{\td \varphi_{12}}{2\pi}\, \cos 2\varphi_{12} \overline{J (s_1,t_1) J (s_2,t_2)}
\frac{{\cal P}_\zeta (v_2 k)}{v_2^3}
\frac{{\cal P}_\zeta  (w_{12} k)}{w_{12}^3} 
\bigg[
\frac{{\cal P}_\zeta (u_2 k)}{u_2^3} +
\frac{{\cal P}_\zeta  (u_1 k)}{u_1^3} 
\bigg] \Bigg\}
\label{eq:Omega-C}
\end{align}
and
\begin{align}
\label{eq:Omega-N}  
\Omega_{\rm GW}^{(2)} (T)
&= \frac{F_{\rm NL}^4}{6} 
\prod_{i=1}^3 
\biggl[\int_1^\infty \td t_i \int_{-1}^1 \td s_i\, \frac{16}{(t_i^2-s_i^2)^2}\biggr] 
\Bigg\{\frac{1}{2}\overline{J^2 (s_1,t_1)} 
{\cal P}_\zeta (v_1 v_2 k) 
{\cal P}_\zeta (v_1 u_2 k) 
{\cal P}_\zeta (u_1 v_3 k) 
{\cal P}_\zeta (u_1 u_3 k)
\\
&
+ \int_0^{2\pi} \frac{\td \varphi_{12}}{2\pi}\frac{\td \varphi_{23}}{2\pi}
\cos (2\varphi_{12} )
\overline{J (s_1,t_1) J (s_2,t_2)}
\frac{{\cal P}_\zeta(u_3 k)}{u_3^3} 
\frac{{\cal P}_\zeta (w_{13} k)}{w_{13}^3}
\frac{{\cal P}_\zeta (w_{23} k)}{w_{23}^3}
\bigg[
    \frac{{\cal P}_\zeta (v_3 k)}{v_3^3} 
+   \frac{{\cal P}_\zeta (w_{123} k)}{w_{123}^3}
\bigg]\Bigg\} \ , 
\nonumber
\end{align}
where we defined $s_i=u_i-v_i$, $t_i=u_i+v_i$, and 
\begin{subequations}
\begin{eqnarray}
    y_{ij}&=&\frac{\cos\varphi_{ij}}{4}\sqrt{(t_i^2-1)(t_j^2-1)(1-s_i^2)(1-s_j^2)}+\frac{1}{4}(1-s_it_i)(1-s_jt_j)\ , \\
    w_{ij}&=&\sqrt{v_i^2+v_j^2-y_{ij}}\, , 
    \qquad 
    w_{123}=\sqrt{v_1^2+v_2^2+v_3^2+y_{12}-y_{13}-y_{23}}\ .
\end{eqnarray}
\end{subequations}
The time-averaged integrated transfer functions $J(u,v)$ were derived in Refs.~\cite{Espinosa:2018eve,Kohri:2018awv,Atal:2021jyo,Adshead:2021hnm,Li:2023qua}, and are
\begin{align}\label{eq:J-ave-12}
& \overline{ J (s_i,t_i)J (s_j,t_j) }
=
\frac{9}{8}\frac{\left(t_i^2-1\right) \left(t_j^2-1\right) \left(1-s_i^2\right) \left(1-s_j^2\right) \left(t_i^2+s_i^2-6\right) \left(t_j^2+s_j^2-6\right) }{\left(t_i^2-s_i^2\right)^3 \left(t_j^2-s_j^2\right)^3}
\nonumber\\
&
\Bigg[
\left( 
    \left(t_i^2+s_i^2-6\right) 
    \ln \left| \frac{t_i^2-3}{3-s_i^2}\right| 
-   2\left(t_i^2-s_i^2\right)
\right) \left(
    \left(t_j^2+s_j^2-6\right) 
    \ln \left| \frac{t_j^2-3}{3-s_j^2}\right|
-   2\left(t_j^2-s_j^2\right)
\right)
\nonumber\\
&
+   \pi^2\Theta\left(t_i-\sqrt{3}\right) \Theta\left(t_j-\sqrt{3}\right) \left(t_i^2+s_i^2-6\right) \left(t_j^2+s_j^2-6\right)
\Bigg]\ .
\end{align}
Assuming a log-normal curvature power spectrum of the form \eqref{eq:PLN}, as done in Section~\ref{sec:SIGWth}, we can compute the different contributions to the SGWB, factoring out the overall scaling with the spectral amplitude $A$ and the non-linear parameter $F_\text{NL}$, as shown in Fig.~\ref{fig:SIGW-fNL_spectra}.

The presence of higher-order corrections to the SIGW spectrum could be inferred from a modulation of the peak close to $k\approx k_*$, if the combination $A F_{\rm NL}^2$ is sufficiently large. 
On the other hand, the shape of the low-frequency tail at $k \ll k_*$ is maintained, i.e., 
\begin{equation}
    \Omega_{\rm GW} (k \ll k_*) \, {\propto}\,  k^3(1 + \tilde A \ln^2(k/\tilde k))\, ,
\end{equation}
where $\tilde  A$ and $\tilde k = \mathcal{O}(k_{*})$ are parameters that depend mildly on the shape of the curvature power spectrum.

\begin{figure}[h!]
    \centering
    \includegraphics[width=0.35\columnwidth]{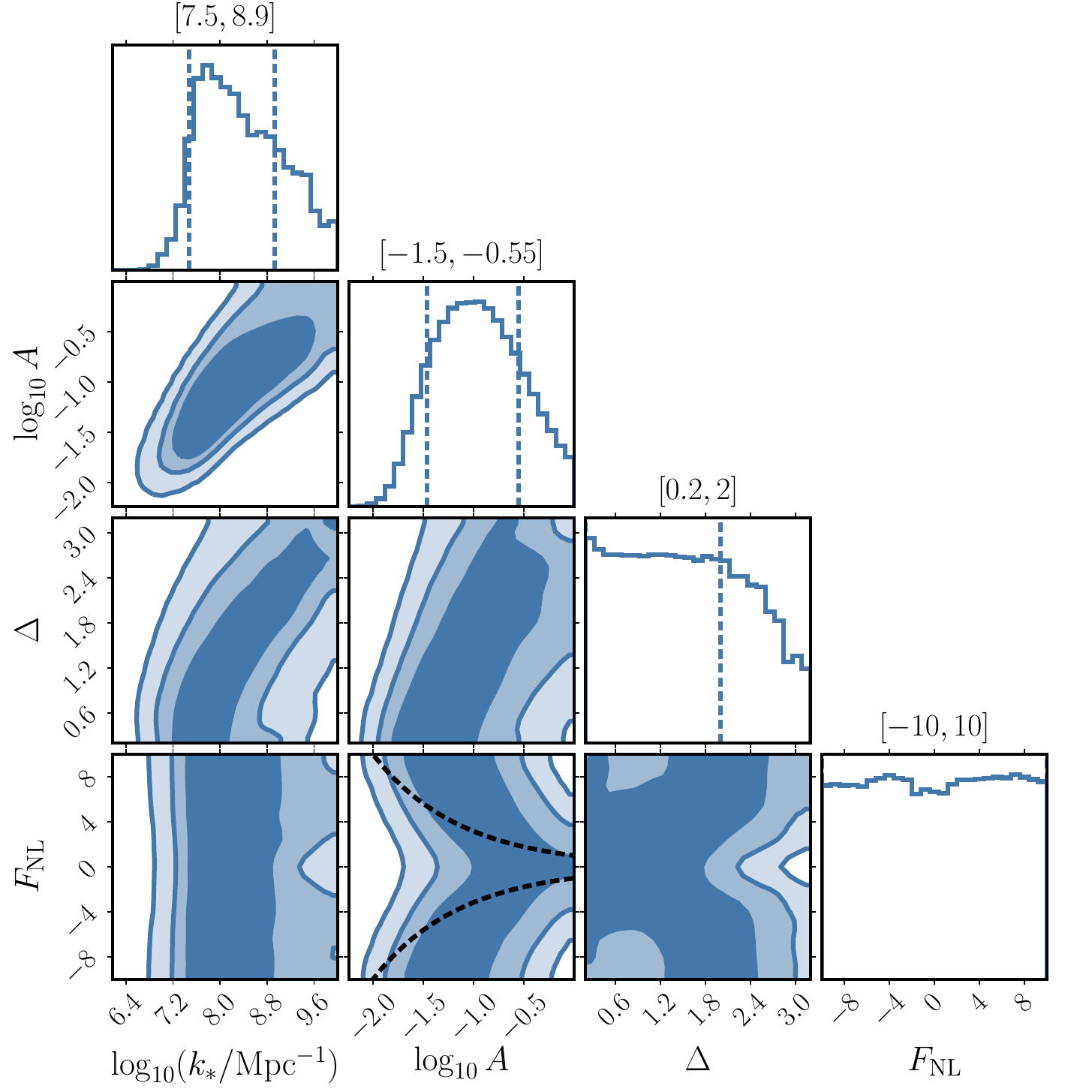}
\includegraphics[width=0.57\columnwidth]{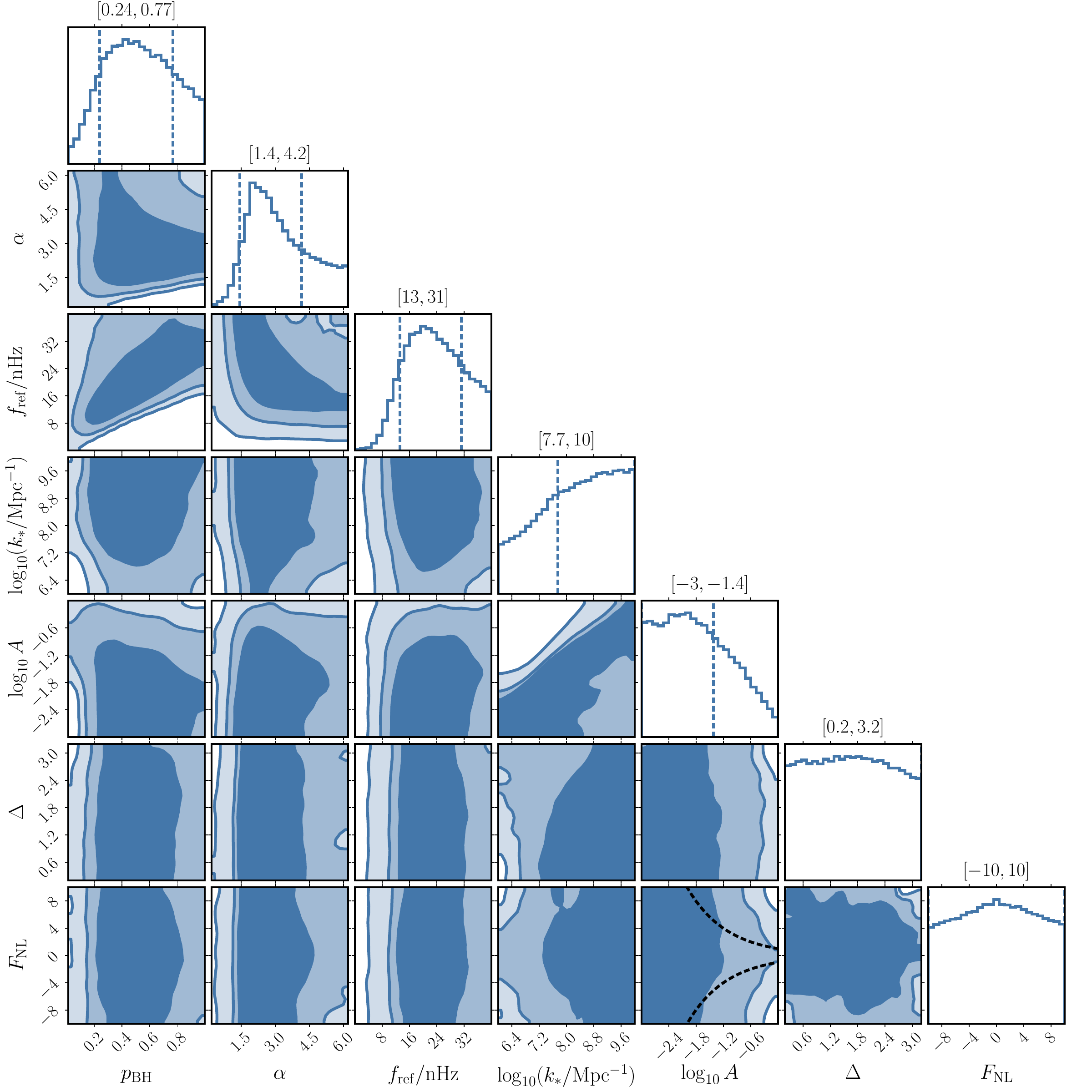}
    \caption{
    Same as Fig.~\ref{fig:SIGWonly_posteriors}
 (left panel) and same as Fig.~\ref{fig:SIGW_posteriors} (right panel) but including the possibility of non-Gaussian curvature perturbations with a non-linear parameter $F_{\rm NL}$.
    }
    \label{fig:SIGW_posteriors-fNL}
\end{figure}

As the PTA observations could only be compatible with the infra-red tail of a SIGW, due to the preference for a positive tilt of the spectrum, we conclude that higher-order corrections do not affect significantly the results presented in the main text, while, in full generality, current GW observations are not able to break the degeneracy in the $(A,F_{\rm NL}^2)$ plane. 
As a by-product of our analysis, at odds with the claim in Refs.~\cite{Liu:2023ymk,Figueroa:2023zhu}, we conclude that current data are not able to constrain the presence of non-Gaussianities at PTA scales, beyond the reach of large-scale structure and CMB probes. This is because large values of $F_{\rm NL}$ remain possible, provided the curvature power spectral amplitude is sufficiently small. 
This is confirmed by the posterior distributions shown in Fig.~\ref{fig:SIGW_posteriors-fNL}, where we report results for non-Gaussian SIGW-only fits (left panel) and for non-Gaussian SIGWs together with SMBH binaries (right panel). 
In both panels, the posterior distribution for $F_{\rm NL}$ is flat over the entire prior range.

\bibliographystyle{apsrev4-1}
\bibliography{refs}

\end{document}